\documentclass[11pt,letterpaper]{article}

\usepackage{amsmath,amssymb,amsthm,amsfonts,amstext,upgreek}
\usepackage{etoolbox}
\usepackage{lscape}
\usepackage{mathtools}
\usepackage{natbib}
\usepackage{bm}
\usepackage{cancel}
\usepackage{tabularx}
\usepackage{adjustbox}	
\usepackage{longtable}
\usepackage{comment}
\usepackage[flushleft]{threeparttable}
\appto\TPTnoteSettings{\linespread{1}\footnotesize} 
\makeatletter 
\g@addto@macro\TPT@defaults{\linespread{1}\footnotesize}
\makeatother
\usepackage{booktabs}
\usepackage{siunitx}
\usepackage{color}
\usepackage{graphicx}
\usepackage{caption}
\usepackage{subcaption}
\usepackage{mathrsfs}
\usepackage{setspace}
\usepackage{rotating}
\usepackage{multirow}
\usepackage{thrmappendix}
\usepackage{alltt}
\usepackage{amsmath}
\usepackage{accents}
\usepackage{mathtools,dsfont}
\usepackage{hyperref}
\definecolor{blue(pigment)}{rgb}{0.2, 0.2, 0.6}
\hypersetup{
	colorlinks=true,
	linkcolor=blue,
	filecolor=magenta,
	urlcolor=cyan,
	citecolor=blue(pigment),
}
\usepackage[capitalize,noabbrev]{cleveref}
\usepackage{array}
\usepackage[inline,shortlabels]{enumitem}

\newtoggle{SUPPLEMENTAL}\toggletrue{SUPPLEMENTAL}
\togglefalse{SUPPLEMENTAL} 

\newtoggle{BLINDED}\toggletrue{BLINDED}
\togglefalse{BLINDED} 

\newcommand{\nn}{\text{norm}}
\newcommand{\ee}{\text{extr}}

\theoremstyle{plain}
\newtheorem{assumption}{Assumption}

\newtheorem{theorem}{Theorem}[]

\newtheorem{proposition}[theorem]{Proposition}

\theoremstyle{definition}
\newtheorem{remark}{Remark}[]

\newcommand{\KK}{{\rm K}}

\renewcommand{\hat}{\widehat}
\DeclareMathOperator{\op}{op}

\crefformat{footnote}{#2\footnotemark[#1]#3} 
\crefname{conjecture}{Conjecture}{Conjectures}
\crefname{section}{Section}{Sections}
\crefname{subsection}{Section}{Sections}
\crefname{subsubsection}{Section}{Sections}
\Crefname{conjecture}{Conjecture}{Conjectures}
\Crefname{section}{Section}{Sections}
\Crefname{subsection}{Section}{Sections}
\Crefname{subsubsection}{Section}{Sections}
\crefname{appendix}{Appendix}{Appendices}
\crefname{subappendix}{Appendix}{Appendices}
\crefname{subsubappendix}{Appendix}{Appendices}
\Crefname{appendix}{Appendix}{Appendices}
\Crefname{subappendix}{Appendix}{Appendices}
\Crefname{subsubappendix}{Appendix}{Appendices}
\crefname{equation}{}{}
\Crefname{equation}{Equation}{Equations}
\crefformat{enumi}{(#2#1#3)}
\crefrangeformat{enumi}{(#3#1#4)\crefrangeconjunction(#5#2#6)}
\crefmultiformat{enumi}{(#2#1#3)}{ and~(#2#1#3)}{, (#2#1#3)}{ and~(#2#1#3)}
\Crefformat{enumi}{Part (#2#1#3)}
\Crefrangeformat{enumi}{Parts (#3#1#4)\crefrangeconjunction(#5#2#6)}
\Crefmultiformat{enumi}{Parts (#2#1#3)}{ and~(#2#1#3)}{, (#2#1#3)}{ and~(#2#1#3)}
\crefname{assumption}{}{}
\Crefname{assumption}{Assumption}{Assumptions}
\newcommand{\crefrangeconjunction}{--}

\DeclareGraphicsExtensions{.pdf,.png}
\graphicspath{ {./Figures/} }

\DeclareMathOperator{\HS}{HS}

\DeclareMathOperator{\sgn}{sgn}

\newcommand{\ZZ}{\mathrm{Z}}
\newcommand{\zz}{z}

\let\originalleft\left
\let\originalright\right
\renewcommand{\left}{\mathopen{}\mathclose\bgroup\originalleft}
\renewcommand{\right}{\aftergroup\egroup\originalright}
\newcommand{\temden}{\phi}

\renewenvironment{thebibliography}[1]{%
	\begin{oldthebibliography}{#1}%
		\setlength{\parskip}{0ex}%
		\setlength{\itemsep}{0ex}%
	}%
	{%
	\end{oldthebibliography}%
}

\allowdisplaybreaks[3]

\renewcommand{\thefootnote}{\fnsymbol{footnote}}
\renewcommand{\hat}{\widehat}
\title{Nonlinear Temperature Sensitivity of Residential Electricity Demand: Evidence from a Distributional Regression Approach\thanks{We deeply appreciate Hyunjin Yang (KEPCO) and Sungro Lee (KOGAS) for generously providing residential electricity demand and local temperature data in the Republic of Korea. Kyungsik Nam's research was supported by the Hankuk University of Foreign Studies Research Fund of 2025.}}
\author{Kyungsik Nam\thanks{Assistant Professor, Division of Climate Change, Hankuk University of Foreign Studies, Yongin-si, Gyeonggi-do, 17035, Republic of Korea. E-mail: \texttt{ksnam@hufs.ac.kr}}, \,\, Won-Ki Seo\thanks{Senior Lecturer, School of Economics, University of Sydney, Camperdown, NSW 2006, Australia. E-mail: \texttt{won-ki.seo@sydney.edu.au}}}

\begin{document}
	\maketitle
	
	\begin{abstract}
		\noindent We estimate the temperature sensitivity of residential electricity demand during extreme temperature events using the distribution-to-scalar regression model. Rather than relying on simple averages or individual quantile statistics of raw temperature data, we construct distributional summaries, such as probability density, hazard rate, and quantile functions, to retain a more comprehensive representation of temperature variation. This approach not only utilizes richer information from the underlying temperature distribution but also enables the examination of extreme temperature effects that conventional models fail to capture. Additionally, recognizing that distribution functions are typically estimated from limited discrete observations and may be subject to measurement errors, our econometric framework explicitly addresses this issue. Empirical findings from the hazard-to-demand model indicate that residential electricity demand exhibits a stronger nonlinear response to cold waves than to heat waves, while heat wave shocks demonstrate a more pronounced incremental effect. Moreover, the temperature quantile-to-demand model produces largely insignificant demand response estimates, attributed to the offsetting influence of two counteracting forces.
	\end{abstract}
	\vspace{0.4\baselineskip}
	\vfill
	
	 \bigskip
	
	
	\setcounter{footnote}{0}
	\renewcommand{\thefootnote}{\arabic{footnote}}
	\pagenumbering{arabic}
	\pagestyle{myheadings}
	\setcounter{page}{0} \thispagestyle{empty}
	
	\baselineskip19.5pt
	\pagenumbering{arabic}
	\newpage
	\section{Introduction}\label{Sec_Intro}
	\noindent Accurately estimating electricity demand during extreme weather events is vital for ensuring energy security and enhancing grid reliability. For energy practitioners and policymakers, precise demand estimation is crucial for balancing supply and demand, preventing power outages, and ensuring that the energy infrastructure can withstand increased loads during such critical periods. As climate change drives more frequent and severe weather events, the importance of precise demand estimation has grown significantly, making it a cornerstone of sustainable energy management. 
	
	\indent The relationship between temperature and electricity demand is inherently nonlinear, as demand typically rises sharply during hot or cold conditions. Traditional approaches to modeling this relationship rely on nonlinear temperature variables such as Heating Degree Days (HDD) and Cooling Degree Days (CDD). However, these methods have notable limitations, including the need to arbitrarily define threshold temperatures. To address this limitation, nonlinear econometric models or density-based approaches that incorporate the full temperature distribution as an explanatory variable have been investigated. These methodologies effectively capture the seasonal dynamics of electricity demand while allowing the threshold temperature to be estimated endogenously within the model.
	
	\indent Nonetheless, unexpected spikes in electricity demand during extreme temperature events present crucial challenges for grid stability, necessitating more advanced econometric modeling techniques. Residential electricity demand, in particular, exhibits heightened sensitivity to temperature extremes, increasing unpredictably under such conditions. Existing nonlinear and density-based econometric approaches often fail to adequately capture these nonlinear demand responses, leading to significant underestimation during critical periods. As climate change amplifies temperature extremes, reassessing and refining demand estimation methodologies is essential for ensuring grid stability and optimizing operational efficiency.
	
	\indent Beyond the inherent challenges of modeling nonlinear temperature-demand relationships, a significant source of error arises from measurement and estimation inaccuracies. In particular, constructing a representative temperature regressor from high-frequency or regional raw data introduces estimation errors, leading to biased and inconsistent demand response estimates. Additionally, the density-based approach assumes that nonparametrically estimated temperature distributions represent true distributions. This assumption further amplifies estimation bias, as errors in the temperature distribution can distort demand response projections. These errors are particularly problematic during extreme temperature events, where precise demand forecasting is crucial for ensuring energy security and grid stability.
	
	\indent Building on the foundational work of \cite{engle1986semiparametric}, a substantial body of literature has applied the cointegration framework for analyzing electricity demand, with noticeable contributions from \cite{engle1989merging}, \cite{silk1997short}, \cite{beenstock1999demand}, \cite{chang2014time}, \cite{chang2016disentangling}, and \cite{chang2021forecasting}, among others. Subsequently, the literature has focused on incorporating nonlinear temperature variables, such as HDD and CDD, or employing nonlinear econometric methods to evaluate the impact of temperature fluctuations on electricity demand, including \cite{sailor1997sensitivity}, \cite{henley1997non}, \cite{henley1998residential}, \cite{valor2001daily}, \cite{pardo2002temperature}, and \cite{apadula2012relationships}. In particular, smooth transition, panel threshold, and switching regression models have been employed to capture nonlinear demand responses (\citealp{moral2005modelling}; \citealp{bessec2008non}). The nonlinear relationship between electricity demand and temperature has been further explored through various advanced methodologies. Notably, artificial neural networks have been applied to improve load forecasting accuracy (\citealp{teixeira2017}; \citealp{caro2020}; \citealp{sharma2020}), while machine learning approaches have been employed to enhance short-run demand predictions (\citealp{al2019short}; \citealp{sultana2022}). 
	
	\indent Rather than analyzing the overall effects of weather variables, recent literature has increasingly focused on the impact of extreme temperatures on peak electricity demand. Early studies employed seasonal autoregressive moving average models with conditional heteroskedasticity to model fluctuations in peak demand (\citealp{sigauke2011prediction}; \citealp{rallapalli2012forecasting}). Later research incorporated tail-quantile estimation and extreme value theory to improve the accuracy of predicting extreme peak demand (\citealp{sigauke2013extreme}; \citealp{chan2015extreme}; \citealp{sigauke2017modelling}; \citealp{sigauke2020modelling}). More recently, density-based functional approaches have emerged to analyze these extreme effects across sectors, providing deeper insights into the nonlinearity between temperature and electricity demand (\citealp{chang2016new}; \citealp{miller2022modeling}).
	
	\indent Compared to the existing literature, we investigate distribution-to-scalar regression models that effectively account for the nonlinear response of residential electricity demand, even in the presence of measurement errors and extreme temperature events—an aspect that has not been comprehensively explored. By addressing the complexities introduced by measurement errors, our approach provides a more robust estimation, setting it apart from traditional techniques that may overlook these critical factors. Furthermore, our approach differs significantly from existing studies by focusing on the nonlinear response of residential electricity demand to changes in temperature distribution, rather than just temperature levels. This allows for a more comprehensive analysis, capturing the broader distributional effects of temperature fluctuations, particularly during extreme temperature events.
	
	\indent Accordingly, we incorporate both density and hazard rate functions (as distributional summaries of temperature), each offering distinct advantages, to construct a comprehensive analytical framework. The density-based approach provides direct interpretability within a compositional data framework, facilitating a clear assessment of temperature's impact on electricity demand across the entire temperature distribution. In contrast, the hazard rate is particularly well-suited for analyzing extreme temperature events, as it incorporates the conditional probability of such occurrences to some extent, thereby enhancing the precision of demand estimations under extreme conditions. 
	To ensure that distributional information is properly incorporated into our econometric framework, we will consider relevant transformations of the temperature densities and hazard rates  (see Section \ref{sec_method_1}). 
	For comparison, we additionally consider the quantile-to-demand model, which is not only a popular approach for utilizing distributional information (see, e.g., \citealp{yang2020quantile}) but also aligns conceptually with density- or hazard-based models. 

	\indent The remainder of this paper is structured as follows. Section \ref{sec_method} outlines the econometric methodology used to estimate the nonlinear response of residential electricity demand while addressing measurement errors. Section \ref{Sec_Data} describes the residential electricity demand and local temperature data from the Republic of Korea. Section \ref{Sec_Empirics} presents the empirical temperature response functions based on density-, hazard-, and quantile-based approaches under hypothetical scenarios and two historical extreme temperature events. This section also includes a discussion of the estimation results derived from these three distributional predictors. Finally, Section \ref{Sec_conclude} provides concluding remarks, with proofs provided in the Appendix.
	
	\section{Econometric methodology}\label{sec_method}
	\subsection{Distributional predictors} 
	\label{sec_method_1}
	We let $y_t$ represent electricity demand, a practical measurement of which for our empirical study is detailed in Section \ref{Sec_Data}, and let $X_t^\circ$ be a function that summarizes the distributional properties of temperature. A possible (and popular) candidate for $X_t^\circ$ may be the probability density function (PDF) of temperature, denoted hereafter $\temden_t$. However, as well documented in the literature, directly using the PDF as a predictor is not advisable in applications of statistical methods developed within a standard Hilbert space setting for functional data analysis. As proposed by \cite{petersen2016}, we thus consider appropriate transformations of $\temden_t$. 
	There are various potential choices for $X_t^\circ$, but in this paper, we mainly focus on the following: (i) the Centered-Log-Ratio (CLR) transformation and (ii) the Log-Hazard-Rate (LHR) transformation of $\temden_t$. Assuming that the PDF of temperature, $\temden_t$, is supported on $[a,b]$, the CLR transformation $X_t^\circ$ is defined as follows: 
	\begin{equation}\label{eqclr}
		X_t^\circ(r) = \log \temden_t(r) - \int_{a}^b \log \temden_t (s) ds, \quad r\in [a,b]. 
	\end{equation}
	Under some mathematical conditions on the underlying PDF $\temden_t$, the transformation from $\temden_t$ to its CLR not only embeds the probability densities on $[a,b]$ into a linear subspace of the $L^2[a,b]$-Hilbert space (the Hilbert space of square-integrable functions on $[a,b]$) but also is invertible; see e.g., \cite{Egozcue2006} and \cite{Boogaart2014}. It is known that the inverse CLR transformation is given by $X_t^\circ (r) \mapsto \exp(X_t^\circ (r)) \big/ \int_{a}^b  \exp(X_t^\circ (r)) dr$ for $r \in [a,b]$, which recovers $\temden_t$. The CLR transformation of $\temden_t$ has been considered a standard choice in models involving distributional functional data, as discussed in several recent articles to be mentioned shortly.
	
	As an alternative, we also consider the LHR transformation of $\temden_t$, which is given as follows: for some small positive $\delta>0$,
	\begin{equation}\label{eqlhr}
		X_t^\circ(r) = \log \left(\frac{\temden_t(r)}{1-\Phi_t(r)}\right) , \quad r\in [a,b-\delta],
	\end{equation}
	where $\Phi_t(s)$ is the cumulative distribution function (CDF) given by $\Phi_t(s)=\int_{0}^s \phi_t(r)dr$. Note that $h_t(r)=\phi_t(r)/(1-\Phi_t(r))$ is the so-called hazard rate  of $\phi_t$ and for this to be well defined, we need to restrict the support into $[a,b-\delta]$ for some small positive $\delta$. It is known that the LHR is also invertible transformation of $\phi_t$ into a linear space under appropriate conditions (see \citealp{petersen2016}). 
	
	It appears to be more common to consider the CLR transformation or its similar alternatives in the literature on density-valued functional data (see, e.g., \citealp{kokoszka2019forecasting, SEO2019, seong2021functional}), and this is partly a reason that we also consider it in the present paper. On the other hand, cases using the LHR transformation seem scarce in the literature, to the best of the authors' knowledge, except for \cite{petersen2016}, who proposed the LHR as a way to avoid potential issues when directly using the PDF. 
	As a slight and obvious modification of the LHR, we define the following, obtained by replacing the hazard rate ($\phi_t(r)/(1-\Phi_t(r))$) with the reversed hazard rate (see \citealp{block1998reversed}), defined by $\phi_t(r)/\Phi_t(r)$, as follows:
	\begin{equation}\label{eqrlhr}
		X_t^\circ(r) = \log \left(\frac{\phi_t(r)}{\Phi_t(r)}\right), \quad r\in [a+\delta,b].
	\end{equation}
	We hereafter call the above as the Log-Reversed-Hazard-Rate (LRHR) transformation.
	

	In the sequel, we will consider a functional linear model connecting the electricity demand $y_t$ with the distributional predictor $X_t^\circ$. It is worth noting that while the CLR, LHR, and LRHR transformations characterize the temperature distribution, they differ in how they summarize distributional information. In the CLR model, the distributional predictor is fully characterized by the PDF $\phi_t(s)$, which can be approximated as the average change of the distribution function of temperature on a small interval $[s,s+\Delta)$ for small $\Delta>0$, i.e,  $\phi_t(s)\approx \Delta^{-1}\mathbb{P}(s \leq S_t < s+\Delta)$, where $S_t$ denote the random variable of temperature level at time $t$. On the other hand, the hazard (resp.\ reversed) rate may be approximately understood as ${\Delta}^{-1}{\mathbb{P}(s \leq S_t < s+\Delta | S_t \geq s)} = {\Delta}^{-1}{\mathbb{P}(s \leq S_t < s+\Delta)}/{\mathbb{P}(s \geq S_t)}$  (resp.\ $\Delta^{-1}{\mathbb{P}(s-\Delta < S_t \leq s | S_t\leq s)}={\Delta}^{-1}{\mathbb{P}(s \leq S_t < s+\Delta)}/{\mathbb{P}(S_t\leq s)}$). For any $s$,  ${\mathbb{P}(S_t \geq s)}$ (resp.\ ${\mathbb{P}(S_t\leq s)}$) indicates the exposure probability of the temperature being higher (lower) than $s$ at time $t$, and thus the hazard (resp.\ reversed hazard) rate, which we consider, may be understood as the PDF weighted by this exposure probability. If we consider any small density shock, which is particularly concentrated on the upper (lower) tail of the temperature PDF and hence may have a significant impact on the electricity demand despite its small magnitude,\footnote{We are mainly interested in the effect of extreme temperature events on electricity demand, and such events are expected to happen with low probability; thus those will only make small changes in the PDF in the lower or upper tails.} then the LHR (resp.\ LRHR) will be much more responsive to this change than the PDF. It is thus expected that the LHR (resp.\ LRHR) will have a higher tendency to comove with electricity demand, particularly during extremely high (resp.\ low) temperature events, and hence, the model involving the LHR (resp.\ LRHR) may be more suitable for capturing and explaining the effect of extreme temperature events at high (resp.\ low) quantiles on electricity demand.
	
	In the literature on density-valued functional data, it seems to be popular to regard the quantile function corresponding to $\phi_t$ as a distributional summary (see e.g., \citealp{yang2020quantile}) and use it for statistical inference. We also briefly discuss this choice of distributional predictor in Section \ref{sec_more}.

\subsection{Measurement errors}
In practice, $\phi_t$ and $\Phi_t$ cannot directly be observed, and thus the distributional predictor $X_t^\circ$, considered in Section \ref{sec_method_1} is not observable. For empirical analysis, it must be constructed from the raw temperature data. For example, in the CLR case, we may replace $\phi_t$ or $\log \phi_t$ with its relevant nonparametric estimate in \eqref{eqclr} (as in \citealp{SEO2019}). However, this replacement necessarily introduces estimation/smoothing errors, and in this case, it is well known that standard estimation methods for functional linear models may lead to inconsistent estimation of the model; see \citet[Section 7]{Benatia2017}, \cite{Chen_et_al_2020}, and \cite{seong2021functional} for a more detailed discussion.  Section \ref{sec_cons_dist} details how the distributional predictors used in our empirical analysis are constructed. 

Subsequently, we discuss econometric methods for studying the distributional impact of temperature on electricity demand that are robust to these estimation/smoothing errors, which naturally exist, in the distributional predictor.

\subsection{Model and estimator} \label{Sec_Model}
For the subsequent discussion, it will be convenient to introduce some notation. We let $\mathcal H$ denote $L^2[a,b]$ (i.e., the Hilbert space of square-integrable functions defined on $[a,b]$) with inner product $\langle g,h \rangle = \int_{a}^b g(s)h(s)ds$ for and norm $\|g\|=\sqrt{\langle g,g \rangle}$, where $g,h \in \mathcal H$; $\mathcal H$ is a commonly considered Hilbert space in the literature on functional data analysis. The distributional predictor $X_t^\circ$ and its measurement $X_t$,  which are introduced in \eqref{sec_method_1}, are understood as random elements in $\mathcal H$. Essential concepts on $\mathcal H$, relevant to the subsequent discussion, are reviewed in Section \ref{sec_app_prelim}. 

To study temperature sensitivity of electricity demand, we consider the following model: 
\begin{equation} \label{eqreg1}
	y_t = \mu +  f (X_t^{\circ}) + \varepsilon_t,
\end{equation}
where  $\mathbb{E}[ \varepsilon_t] = 0$ and $f:\mathcal H\mapsto \mathbb{R}$ is a linear map, connecting $y_t$ and $X_t^{\circ}$.\footnote{The linear map $f$ can  equivalently be expressed as an integral transformation associated with a kernel function $\mathrm k_f$, as follows: $	f(X_t^{\circ}) = \int_{a}^b \mathcal \mathrm k_f (r) X_t^{\circ}(r)dr$. Obviously, estimation $\mathrm{k}_f$ is identical to estimation of $f$. In this paper, we focus on $f$ but the subsequent results can be rephrased for $k_f$ with only a minor modification.} 
Equation \eqref{eqreg1} corresponds to the Function-to-Scalar Regression (FSR) model, which is widely discussed in the literature. Following the terminology used in conventional linear models, $f$ is typically referred to as the (functional) coefficient or slope parameter. However, in our context, since $f$ captures how the distributional properties of $X_t^\circ$ influence $y_t$, we refer to it as the distributional coefficient hereafter, and we refer to \eqref{eqreg1} as the Distribution-to-Scalar Regression (DSR)  model. 

As discussed in Section \ref{sec_method_1}, in practice, $X_t^\circ$ is not directly observed and has to be estimated or smoothed from discrete samples $\{\phi_t^{\circ} (s_i)\}_{i=1}^{N_t}$ of $\phi_t$, and practitioners often have no choice but to replace $X_t^{\circ}$ with its feasible estimate $X_t$. As earlier pointed out by \citet[Sections 2 and 5.3]{seong2021functional}, this is a common source of endogeneity in the regression model involving functional data, and ignoring such errors may be detrimental to the use of standard estimators of $f$, requiring the exogeneity of the predictor, developed in the literature. Even if more detailed discussion can be found in recent articles (e.g., \citealp{Chen_et_al_2020,seong2021functional}), it will be helpful to illustrate this issue with a concrete example for the subsequent discussion.  

Suppose that $X_t = X_t^{\circ} + e_t$, where $e_t$ represents an additive deviation of $X_t$ from $X_t^{\circ}$ and satisfies $\mathbb{E}[e_t] = 0$. From \eqref{eqreg1}, we obtain the following relationship: \begin{equation} \label{eqreg2} y_t = \mu + f(X_t) + u_t, \quad u_t = f(e_t) + \varepsilon_t, \end{equation} where $\mathbb{E}[u_t] = 0$. Given that $X_t^\circ$ is unobserved, while $X_t$ is observable, practitioners will work with \eqref{eqreg2} rather than \eqref{eqreg1}. In \eqref{eqreg2}, however, $X_t$ and $u_t$ are generally correlated (meaning that $\mathbb{E}[\langle X_t,x \rangle  u_t] \neq 0$ for some $x\in \mathcal H$) and thus $X_t$ is endogenous. 
As is well documented in the aforementioned articles, this endogeneity makes the standard estimators, developed under the exogeneity condition ($\mathbb{E}[\langle X_t,x\rangle u_t]=0$ for every $x\in \mathcal H$), inconsistent and invalidates the inferential methods based on those estimators (see also \citealp{Benatia2017}). 

To estimate the distributional coefficient $f$ from \eqref{eqreg2} and implement valid statistical inference in our DSR model, which inevitably involves measurement errors, we adapt the functional IV approach, considered in \cite{seong2021functional} for Function-to-Function Regression (FFR) models, to our context. To this end, we let $Z_t$ be the functional Instrumental Variable (IV), which satisfies the following: 
\begin{align}\label{eqreg3}
	C_{XZ}(x)&:=\mathbb{E}[\langle X_t-\mathbb{E}[X_t], x \rangle (Z_t-\mathbb{E}[Z_t])]\neq 0 \quad \text{for some $x\in\mathcal H$},\\ 
	C_{Zu}(x)&:=\mathbb{E}[\langle Z_t-\mathbb{E}[Z_t],x\rangle  u_t]=0 \quad \text{for every nonzero $x\in\mathcal H$}. \label{eqreg3a} 
\end{align}
$C_{XZ}$ (resp.\ $C_{Zu}$) defined above is called the (cross-)covariance operator of $X_t$ and $Z_t$ (resp.\ $Z_t$ and $u_t$). The conditions given in \eqref{eqreg3} and \eqref{eqreg3a} consist of only the minimal requirements for the IV $Z_t$ in our DSR model. More conditions for the consistency and (local) asymptotic normality of our estimator will be detailed in Section \ref{sec_app_est} of the Appendix.
Assuming that the measurement errors are not serially correlated and that $X_t$ tends to exhibit time series dependence in the considered empirical model, we may let $Z_t$ be a lagged distributional predictor in our empirical analysis, as in \citet[Section 5.3]{seong2021functional}; this is also a special case of the IV considered in \cite{Chen_et_al_2020}. To obtain our proposed estimator, we need to compute the sample covariance operators $\widehat{C}_{XZ}$ and its adjoint $\widehat{C}_{XZ}^\ast$ defined as follows: for any $x \in \mathcal H$,
\begin{equation}\label{eqcxz}
	\widehat{C}_{XZ}(x) = \frac{1}{T}\sum_{t=1}^{T} \langle X_t-\bar{X}_T,x\rangle (Z_t-\bar{Z}_T), \quad \widehat{C}_{XZ}^\ast(x) = \frac{1}{T}\sum_{t=1}^{T} \langle Z_t-\bar{Z}_T,x \rangle (X_t-\bar{X}_T), 
\end{equation}
where $\bar{X}_T=T^{-1}\sum_{t=1}^T X_t$ and $\bar{Z}_T=T^{-1}\sum_{t=1}^T Z_t$. $\widehat{C}_{XZ}$ and $\widehat{C}_{XZ}^\ast$ are, respectively, the sample counterparts of $C_{XZ}$ and its adjoint $C_{XZ}^\ast$ (defined by $C_{XZ}^\ast(x) = \mathbb{E}[\langle Z_t-\mathbb{E}[Z_t],x \rangle(X_t-\mathbb{E}[X_t])])$.  As is well known in the literature, the composite map $\widehat{C}_{XZ}^\ast \widehat{C}_{XZ}$ allows the eigendecomposition with respect to its nonnegative eigenvalues (hereafter denoted $\{\hat{\lambda}_j^2\}_{j\geq 1}$) and the corresponding eigenvectors (denoted $\{\hat{g}_j\}_{j\geq 1}$). That is, 
\begin{equation*}
	\widehat{C}_{XZ}^\ast \widehat{C}_{XZ} \hat{g}_j =  \hat{\lambda}_j^2 \hat{g}_j, 
\end{equation*}
where $\hat{\lambda}_1^2\geq \hat{\lambda}_2^2 \geq \ldots \geq 0$. The eigenelements of $\widehat{C}_{XZ}^\ast \widehat{C}_{XZ}$ can be computed using the standard functional principal component analysis (FPCA) as in \citet[pp.\ 117-118]{Bosq2000} and \cite{seong2021functional}. 
Then our proposed estimator $\hat{f}$, which is a map from $\mathcal H$ to $\mathbb{R}$, is defined as follows: for any $x \in \mathcal H$,
\begin{align} \label{eqest}
	\hat{f}(x) = T^{-1}\sum_{t=1}^T \sum_{j=1}^{\KK} \hat{\lambda}_j^{-2} \langle \hat{g}_j, x \rangle \langle \widehat{C}_{XZ}\hat{g}_j,Z_t-\bar{Z}_T \rangle  (y_t-\bar{y}_T),
\end{align}
where $\bar{y}_T=T^{-1}\sum_{t=1}^T y_t$ and $\KK$ is sufficiently smaller than $T$; a more detailed and technical requirement on the choice of $\KK$ will be postponed to Section \ref{sec_app_est} of the Appendix. If $\KK$ is determined, it is straightforward to compute the estimator using  $\widehat{C}_{XZ}$ and the sample eigenelements $\{\hat{\lambda}_j^2,\hat{g}_j\}$. The above estimator turns out to be viewed as a sample analogue estimator based on the population equation $C_{Zy}C_{XZ}= f C_{XZ}^\ast C_{XZ}$, which holds when $Z_t$ satisfies \eqref{eqreg3} (see Section \ref{sec_app_est} of the Appendix for more detailed discussion). From this equation, the estimator \eqref{eqest} is obtained by (i) replacing the population operators $C_{Zy}$ and $C_{XZ}$ with their sample counterparts,  $\widehat C_{Zy}$ and $\widehat C_{XZ}$, respectively, and then (ii) employing a regularized inverse of $\widehat{C}_{XZ}^\ast \widehat{C}_{XZ}$, which is necessary to compute the estimator.
Through our study of the asymptotic properties of the proposed estimator,  we establish the consistency of $\hat{f}$ under appropriate assumptions as follows:
\begin{proposition}\label{prop1}
	Under Assumption \ref{assum1} given in Section \ref{sec_app_est}, $\hat{f}$ is a consistent estimator of $f$. 
\end{proposition}
Proposition \ref{prop1} is presented without the mathematical details intentionally; a more rigorous version is provided in the Appendix (see Proposition \ref{prop1a}). The consistency stated in Proposition \ref{prop1} implies that for every $x \in \mathcal{H}$, $\hat{f}(x)$ converges in probability to $f(x)$. 

It is often of interest to practitioners to make inference on $f(\zeta)$, which may be understood as the effect of an additive distributional perturbation $\zeta$ given to $X_t^\circ$ on $y_t$. 
{Specifically, in our empirical analysis, $\zeta$ will be set to various extreme temperature events and we are interested in the effect of these events on $y_t$.} The consistency result in Proposition \ref{prop1} implies that this effect can be estimated by $\hat{f}(\zeta)$. Beyond the consistent estimation, the following result, which can be used for statistical inference on $f(\zeta)$, is established under some appropriate conditions (see Proposition \ref{prop2} in the Appendix):  
\begin{equation}\label{eqlocal}
	\sqrt{\frac{T}{\hat{\sigma}^2_u \widehat{\theta}_{\KK}(\zeta)}}(\hat{f}(\zeta)-f(\zeta)) \to_d N(0,1),
\end{equation}
where   $\hat{\sigma}^2_u = T^{-1} \sum_{t=1}^T \hat{u}_t^2$ with $\hat{u}_t = y_t-\hat{f}(X_t)$ and
\begin{equation} \label{eqtheta}
	\widehat{\theta}_{\KK}(\zeta) 
	=\frac{1}{T} \sum_{t=1}^T \left(\sum_{j=1}^{\KK} \hat{\lambda}_j^{-2} \langle \hat{g}_j, \zeta\rangle \langle \widehat{C}_{XZ}\hat{g}_j, Z_t-\bar{Z}_T\rangle\right)^2.
\end{equation}
The result \eqref{eqlocal} may be understood as the local asymptotic normality of the functional IV estimator, which was earlier obtained by \cite{seong2021functional} in the FFR model.    

It is well known that  $f(\zeta)$ for any $\zeta$ may be understood as the integral transformation 
\begin{equation*}
	f(\zeta)   =   \int_{a}^b \psi_f(r) \zeta (r) dr
\end{equation*}
for some uniquely identified function $\psi_f \in \mathcal H$.   $\psi_f(s)$ is sometimes interpreted as the temperature effect when the distributional predictor $X^\circ$ hypothetically concentrated at a point $s$ (\citealp{chang2014time}). It might be of interest to practitioners to implement statistical inference on $\psi_f(s)$, but in our setup, it is not possible to implement direct statistical inference with a perturbation which is fully concentrated at a point.\footnote{In the standard $L^2[a,b]$-Hilbert space setting, such a function is essentially equivalent to the zero function.} However, it is, instead, possible to consider a weighted average of $\psi_f(s)$ near $s$. For example, if we let $\zeta_s(r) = w_h(|r-s|)/{\int_{a}^{b} w_h(|r-s|)dr}$, then statistical inference on the following quantity is feasible based on the asymptotic normality result \eqref{eqlocal}:
\begin{equation} \label{eqkernel}
	f(\zeta_s) =  \frac{\int_{a}^{b} w_h(|r-s|) \psi_f(r) dr}{\int_{a}^{b} w_h(|r-s|)dr}  \eqqcolon \tilde{\psi}_f(s), 
\end{equation}
where $w_h(|r-s|)$ is the standard kernel function defined near $s$ with bandwidth $h$. The above $\tilde{\psi}_f(s)$ is dependent on the choice of kernel function, but it may be understood as the local weighted average of $\psi_f$ near $s$ for any reasonable choice of the kernel function. We hereafter call $\tilde{\psi}_f(s)$ as the kernel-weighted temperature response function.

\section{Data}\label{Sec_Data}
\subsection{Raw data and temperature densities} \label{Sec_Data_1}

\noindent To investigate the nonlinear temperature sensitivity of electricity demand, we analyze an extensive dataset that includes residential electricity demand (in GWh) and local temperature (in degrees Celsius) readings from the Republic of Korea. The dataset spans from January 1999 to December 2023, with 300 monthly observations. The residential electricity demand data is provided by the Korea Electric Power Corporation (KEPCO), the state-owned utility responsible for electricity sales across the country. Due to the inherent variability in the number of days within each month, we have normalized the raw monthly electricity demand by dividing it by the number of effective days per month. This adjustment produces a metric known as electricity demand per effective day, as discussed in \cite{chang2014time}, which underscores the importance of accounting for temporal differences in energy demand analysis.

\begin{figure}[t]
\begin{center}
\includegraphics[height=0.45\textwidth, width=0.8\textwidth]{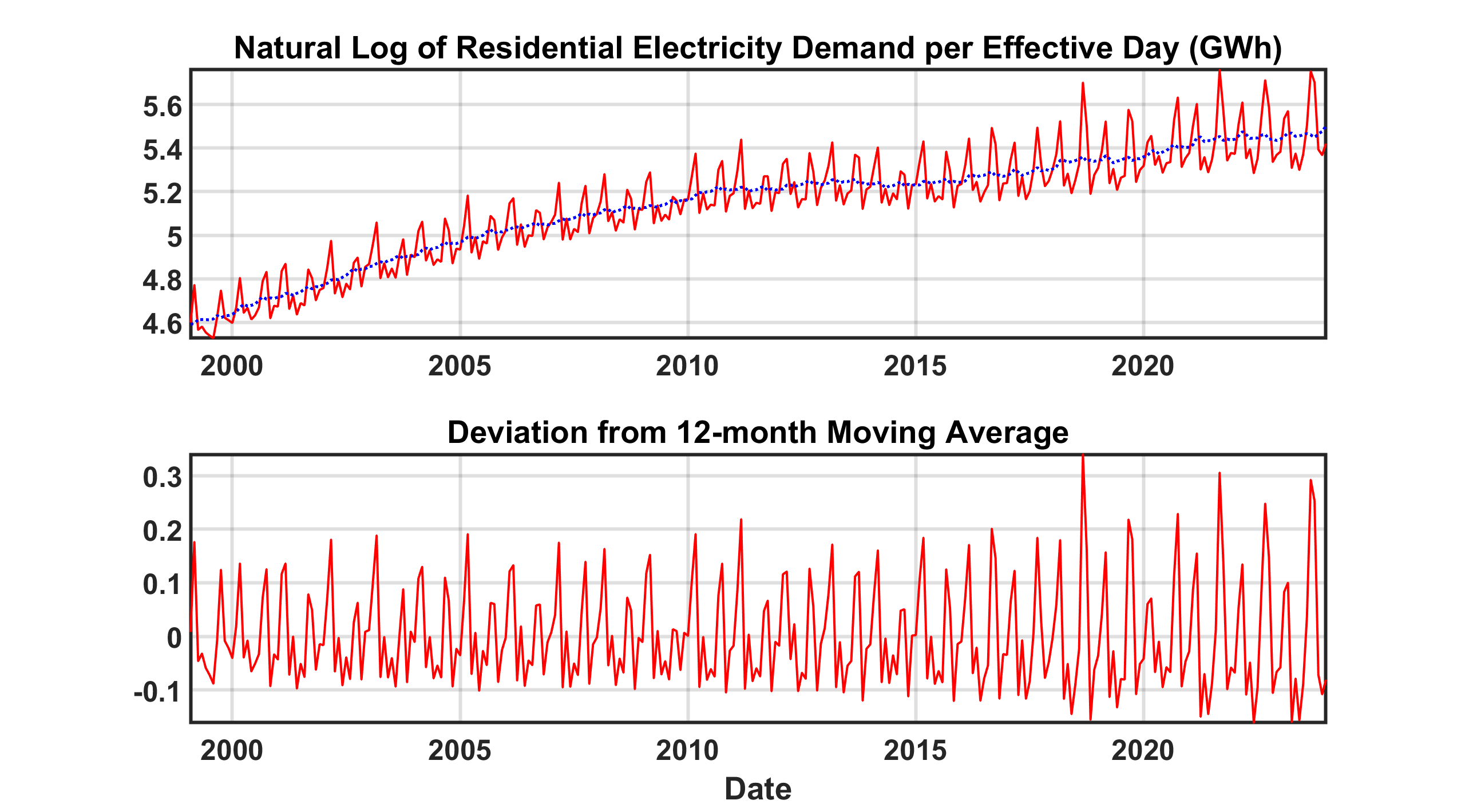}
\caption{Natural log of residential electricity demand per effective day and its 12-month moving average (top), and their deviation process (bottom), from January 1999 to December 2023.}
\label{Fig:Y_Graphs}
\end{center}
\end{figure}

\indent \cite{chang2016new} found no empirical evidence suggesting that non-climate variables significantly influence the demand response to temperature in the residential sector. Building on this finding, and to isolate the short-run fluctuations in residential electricity demand, we calculate the deviation from the 12-month moving average of the natural logarithm of electricity demand per effective day, as illustrated in Figure \ref{Fig:Y_Graphs}. This deviation serves as a proxy for the short-run component of electricity demand, allowing us to focus on the short-run effects of temperature changes while filtering out the long-run effects of income and electricity price. By employing this detrending approach, we ensure that our analysis effectively captures the nonlinear responses of residential electricity demand to temperature fluctuations, independent of other non-climate factors (\citealp{chang2016new}). 

\indent To model the seasonality, we retrieve the hourly temperature data from the KMA National Climate Data Center.\footnote{Downloaded from https://data.kma.go.kr/cmmn/main.do on July 1, 2024.} We then construct the monthly temperature PDF over the common support, $[-20,40]$, using hourly temperature observations from the five largest cities (Seoul, Daejeon, Daegu, Gwangju, and Busan) covering the territory of the Republic of Korea.\footnote{Instead of estimating the range of the common support, we adopt the common support of $[-20,40]$ with a grid difference, $0.1$, as provided in the literature (\citealp{chang2014time}; \citealp{chang2016new}).} To represent the temperature PDF of the Republic of Korea, we calculate the weighted average of the estimated time series for these five PDFs, assigning weights based on the annual proportion of electricity sales in each of the five regions: the Seoul Metropolitan Area (Seoul, Incheon, Gyeonggi), Daejeon, Daegu, Gwangju, and Busan.\footnote{The combined annual electricity consumption of the five regions accounts for approximately 55.3\% of the national total on average during the period from 1999 to 2021. Due to data availability constraints, the proportions for the years 2022 and 2023 have been substituted with the values from 2021.}\footnote{The annual regional consumption data was obtained from the Korea Energy Statistics Information System (KESIS) and can be accessed at \url{https://www.kesis.net/sub/subChart.jsp?report_id=33150&reportType=0} (data downloaded on July 1, 2024).} 

\begin{figure}[t]
\begin{center}
\includegraphics[height=0.35\textwidth, width=0.4\textwidth]{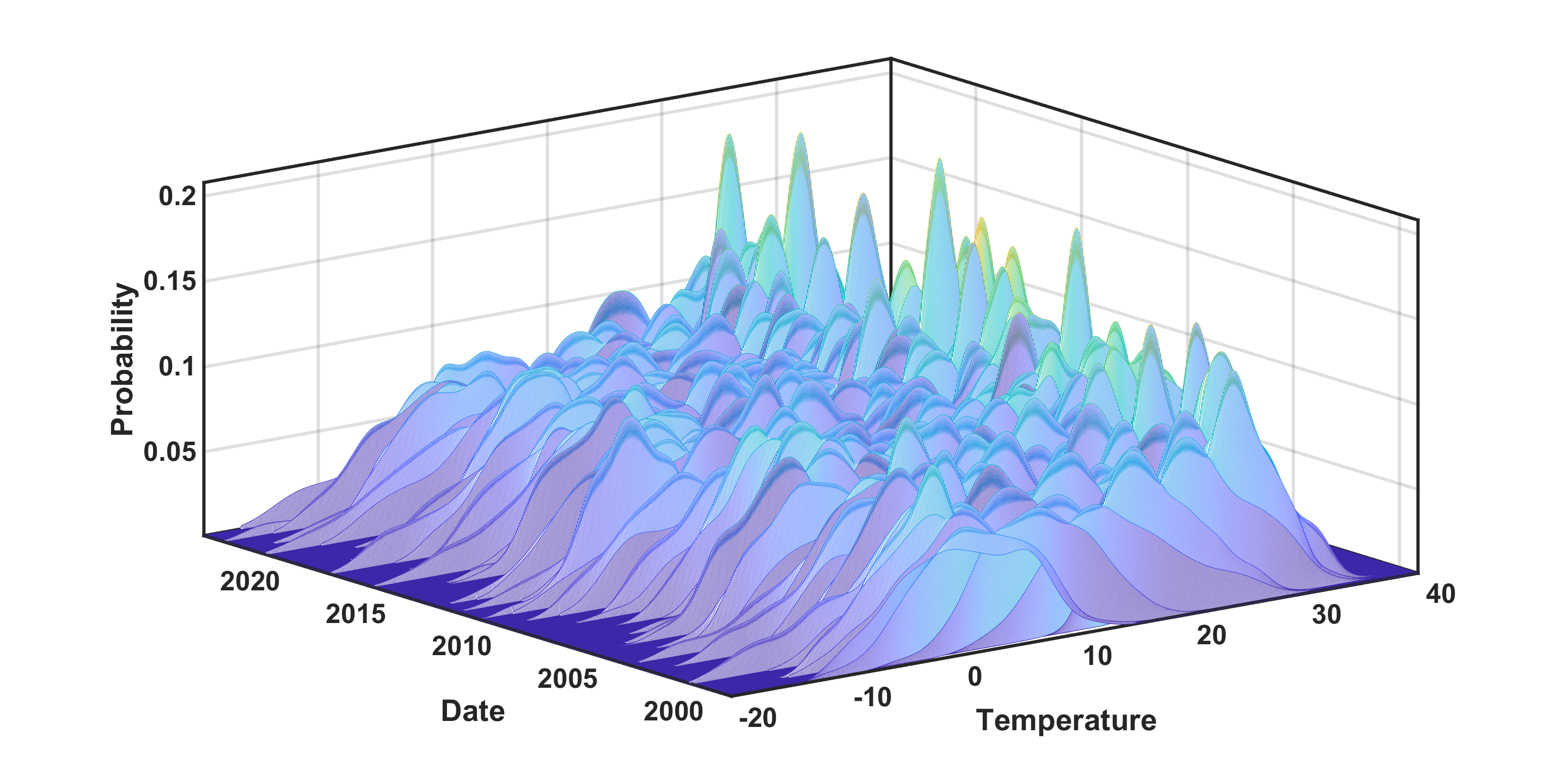}
\includegraphics[height=0.35\textwidth, width=0.55\textwidth]{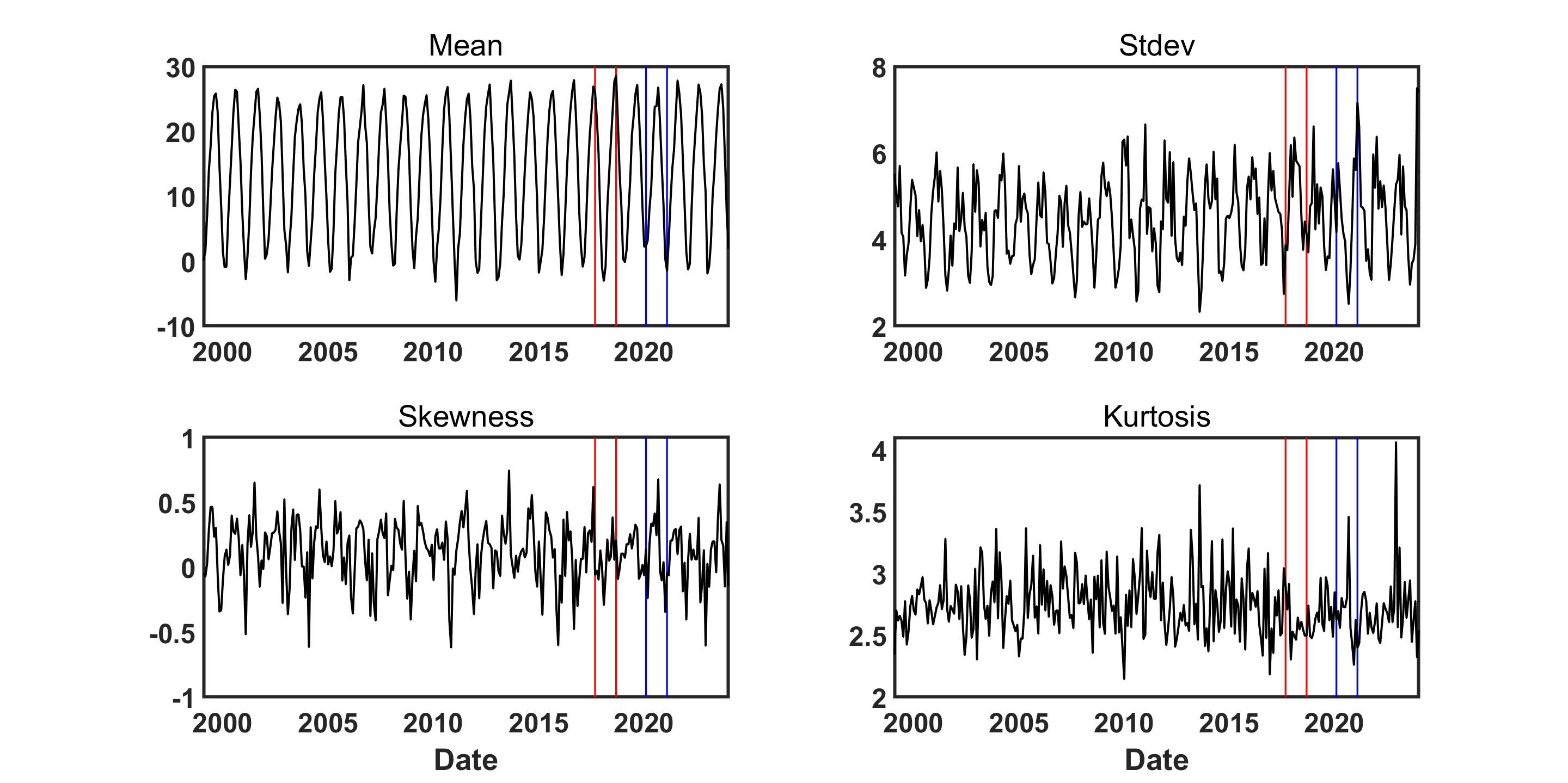}
\caption{Kernel density estimates of the temperature PDF in the Republic of Korea (left) and the dynamics of its descriptive statistics (right). The red lines indicate August 2017 and August 2018, while the blue lines represent January 2020 and January 2021.}
\label{Fig:Quantile_descrip}
\end{center}
\end{figure}

\indent Figure \ref{Fig:Quantile_descrip} illustrates the time series of temperature PDFs, estimated using Gaussian kernel and Silverman's rule of thumb bandwidth, and the dynamics of their descriptive statistics. Note that the generated monthly temperature PDFs exhibit significant heterogeneity, particularly between summer and winter. The time series of monthly temperature distributions in the Republic of Korea from 1999 to 2023 reveals stable seasonal cycles in the mean temperature, reflecting consistent annual patterns. However, the standard deviation indicates increasing variability, particularly post-2010, suggesting a growing spread in temperature values around the mean. The skewness exhibits seasonal fluctuations, with positive skewness during warmer months indicating a higher frequency of extreme high temperatures, and negative skewness during colder months reflecting an increase in extreme low temperatures. However, this skewness does not show a clear trend of increasing asymmetry over time but rather indicates a recurring seasonal pattern. The kurtosis remains relatively stable but experiences significant spikes, particularly after 2010. These spikes suggest periods with more frequent and intense extreme temperature events. Thus, while the skewness reflects consistent seasonal variability, the spikes in kurtosis indicate a rise in temperature extremes, likely influenced by climate change. 	

\subsection{Construction of the distributional predictors} \label{sec_cons_dist}
As discussed in Section \ref{sec_method_1}, distributional predictors in practice need to be estimated from the raw data. For the CLR transformations, we use the estimated densities, denoted hereafter as $\hat{\phi}_t$, obtained in Section \ref{Sec_Data_1} (see Figure \ref{Fig:Quantile_descrip}), and construct $X_t$ as follows:
\begin{equation}\label{eqclr2} X_t(r) = \log \hat{\phi}_t(r) - \int_{-20}^{40} \log \hat{\phi}_t (s) ds.\footnote{As is common in the literature (see e.g., Section 5.2, \citealp{seoshang22}), a small value of $10^{-3}$ is added to the entire temperature PDF $\hat{\phi}_t$ to implement the log transformation over the common support range of $[-20,40]$.}  
\end{equation} 
For the model using the LRH and LRHR predictors, $X_t$ can be constructed as
\begin{equation}\label{eqlhr2} 
X_t(r) = \log \left(\frac{\hat{\phi}_t(r)}{1-\widehat{\Phi}_t(r)}\right) \quad \text{and} \quad  X_t(r) = \log \left(\frac{\hat{\phi}_t(r)}{\widehat{\Phi}_t(r)}\right), 
\end{equation}
where $\widehat{\Phi}_t(r) = \int_{0}^{r} \hat{\phi}_t(s)ds$. However, in practical computation for the LHR predictor, instead of $1-\widehat{\Phi}_t(s)$, we take the maximum of a small positive constant ($10^{-3}$ in the subsequent analysis) and $1-\widehat{\Phi}_t(s)$ in \eqref{eqlhr2}, which is done to avoid the case where $1-\widehat{\Phi}_t(s)$ equals zero, and hence the hazard rate is undefined. The LRHR predictor is similarly constructed by taking the maximum of a small positive constant and $\widehat{\Phi}_t(s)$.\footnote{As an alternative, we also considered adding 0.001 to $1-\widehat{\Phi}_t(s)$ or $\widehat{\Phi}_t(s)$, but we found that the estimation results change little.} Note that this convenient method of constructing the LHR and LRHR results in the constructed predictors not perfectly matching their theoretical counterparts (\eqref{eqlhr} and \eqref{eqrlhr}). However, for our analysis, we only require that the constructed predictors serve as reasonable proxies, and thus, this does not invalidate the empirical results that will be discussed.


\bigskip
\section{Empirical Investigation}\label{Sec_Empirics}
\noindent In this section, we estimate the DSR model \eqref{eqreg2} with potential endogeneity, using the distributional predictors described in \eqref{eqclr2} and \eqref{eqlhr2}. 
For the computation of our estimator with finite samples, we represent the distributional predictor using 80 orthonormal Fourier basis functions on the support.\footnote{Moderate changes in the number of basis functions result in only minor numerical differences in the estimation results.} Noting that measurement errors mostly arise from smoothing \( X_t \) from its discrete realizations, we assume that these errors are serially uncorrelated. Under this assumption, and given that the sequence of \( X_t^\circ \) (and also \( X_t \)) is serially correlated, the lagged variable \( X_{t-\kappa} \) for \( \kappa \geq 1 \) may serve as a candidate IV for our empirical analysis. Similar approaches were previously adopted in  \cite{Chen_et_al_2020} and \cite{seong2021functional} concerning dependent functional data. Since our approach relies on a high degree of dependence, where possible, of the IV on the endogenous distributional predictor \( X_t \), as in standard IV methods (see Section \ref{sec_app_est_1} for more details), it is not advisable to use a large value of \( \kappa \), as this could make \( X_{t-\kappa} \) only weakly correlated with \( X_t \). Therefore, we choose to set \( \kappa \) to 1. We believe that this selection allows us to account for potential errors in the density/quantile estimation process or data quality while enhancing the robustness and reliability of our analysis. Lastly, in our estimation procedure,  $\KK$ (see \eqref{eqest}) is set to 3.\footnote{Following the results given in Section \ref{sec_app_est} of the Appendix and Remark \ref{rem1}, we set $\KK=1+\max_{j\geq 1}\{\tilde{\lambda}_j^2 \geq \alpha\}$ for $\alpha = 0.01 \times T^{-0.2}$, where $\tilde{\lambda}_j^2=\lambda_j^2/\sum_{j=1}^\infty \hat{\lambda}_j^2$ and this is to obtain a choice of $\KK$ which is independent of the scale of $X_t$ and $Z_t$; see Section S5 of \cite{seong2021functional}.} 

\indent We hereafter focus on statistical inference regarding the impact of temperature events on electricity demand, given by $f(\zeta)$, where $\zeta$ denotes the distributional shocks applied to the distributional predictor. This quantity is estimated using our proposed estimator, $\hat{f}(\zeta)$, and is accompanied by an asymptotically valid confidence interval constructed based on the results in \eqref{eqlocal}. Taking advantage of the distributional flexibility of our DSR model, we examine hypothetical scenarios and two historical events for the specification of $\zeta$, as detailed in the following section.
\subsection{Functional Shocks for Cold Wave and Heat Wave}\label{Sec_Est_Shocks}

\noindent In this paper, we present two temperature response functions. The first is the kernel-weighted temperature response function (see \eqref{eqkernel}), which serves as a benchmark and is similar to those estimated by \cite{chang2014time} and \cite{chang2016new}. It captures the point-wise temperature sensitivity across the extreme temperature domain. The second is derived from historical extreme temperature events. While the benchmark temperature response function provides intuitive insights into temperature sensitivity by isolating the effects of temperature changes from distributional shifts, the response function based on historical extreme events enhances the practical relevance of extreme temperature fluctuations.

\indent More specifically, the key distinction between these two approaches lies in their treatment of distributional changes. The analysis of historical extreme events explicitly accounts for shifts in the distribution as normal temperature distributions transition toward extreme ones. In contrast, the benchmark temperature response function hypothetically assumes that the temperature stays at a specified level without consideration of the distributional shift that appeared in the past. Although we provide representative cases of extreme temperature events, the specific distributional characteristics of past cold or heat waves may not necessarily recur. Therefore, the difference between the benchmark response function and the date-specific response function can be interpreted as the additional distributional effect unique to that date.

\begin{figure}[t]
\begin{center}
\includegraphics[height=0.3\textwidth, width=0.98\textwidth]{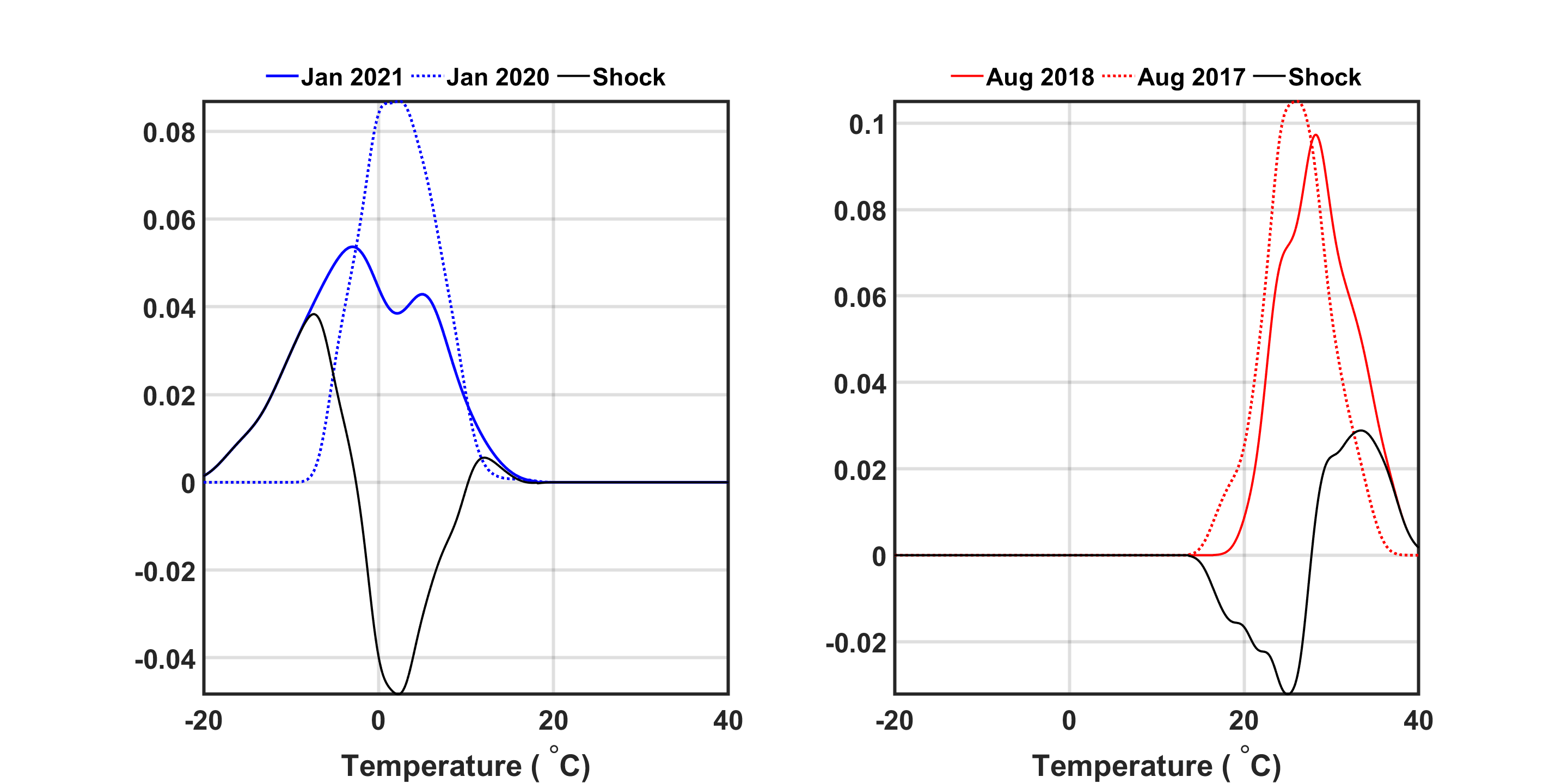}
\caption{Kernel density estimates comparing January 2021 (solid) and January 2020 (dotted) on the left, and August 2018 (solid) and August 2017 (dotted) on the right, with the black lines representing their vertical differences.}
\label{Fig:Func_shock2}
\end{center}
\end{figure}

\indent To estimate the benchmark response function, $\tilde{\psi}_f(s)$ (see \eqref{eqkernel}), we employ Gaussian kernel, which is truncated on the support; more specifically, for each temperature level \( s \), $w_h(|r-s|)$ is defined as follows:
\[
w_h(|r-s|) = \frac{1}{\sqrt{2\pi}} \exp\left\{-0.5\left(\frac{|r-s|}{h}\right)^2\right\} \times {1}\{ -20 \leq s \leq 40\},
\]
where the bandwidth \( h \) is set to span 15°C around \( s \) to align with the range of density increases observed in cold (17.3°C) and heat (12.2°C) wave shocks, as illustrated by the black lines of Figure \ref{Fig:Func_shock2}. 

\indent For the response function based on historical extreme events, we consider two significant extreme weather events that occurred in the Republic of Korea during the sample period.\footnote{See e.g., \url{https://www.wunderground.com/cat6/Hottest-Day-Korean-History} and \url{https://watchers.news/2021/01/11/historic-cold-wave-and-heavy-snow-hit-south-korea/}.} In August 2018, the Republic of Korea experienced one of its most intense heat waves on record, with temperatures soaring to an average of 29.2°C, the highest since record-keeping began in 1907. This extreme heat led to the residential electricity demand of 8,851.04 GWh during that month, underscoring the strain placed on the energy grid by cooling needs. Similarly, January 2021 was marked by a significant cold wave, with temperatures plunging to an average of -6.1°C, well below the seasonal norm. During this period, residential electricity demand also spiked to 7,163.28 GWh as households increased heating usage to cope with the severe cold. 

\indent Figure \ref{Fig:Func_shock2} illustrates the year-over-year changes in monthly temperature PDFs associated with extreme weather events, specifically the heat wave of August 2018 and the cold wave of January 2021. In the left panel of Figure \ref{Fig:Func_shock2}, the blue solid line represents the temperature PDF for January 2021, compared to the blue dashed line for January 2020, with the black line showing the (vertical) difference between the two PDFs, capturing the effect of the cold wave. Similarly, in the right panel of Figure \ref{Fig:Func_shock2}, the red solid line represents the temperature PDF for August 2018, while the red dashed line shows the temperature PDF for August 2017, with the black line indicating the difference between these two PDFs.

\indent The temperature distributions during the 2018 heat wave and the 2021 cold wave exhibit distinct shifts, indicative of extreme weather conditions. The August 2018 distribution is significantly right-skewed, suggesting a higher frequency of exceptionally high temperatures, with a pronounced right tail representing extreme heat events. In contrast, the January 2021 distribution is heavily left-skewed, highlighting the prevalence of unusually low temperatures, with a long left tail representing severe cold conditions. These graphs reveal significant changes in temperature distributions during these historical extreme events, illustrating how both the heat wave and cold wave deviated sharply from the typical conditions of the preceding months.

\begin{figure}[t]
\begin{center}
\includegraphics[height=0.3\textwidth, width=0.54\textwidth]{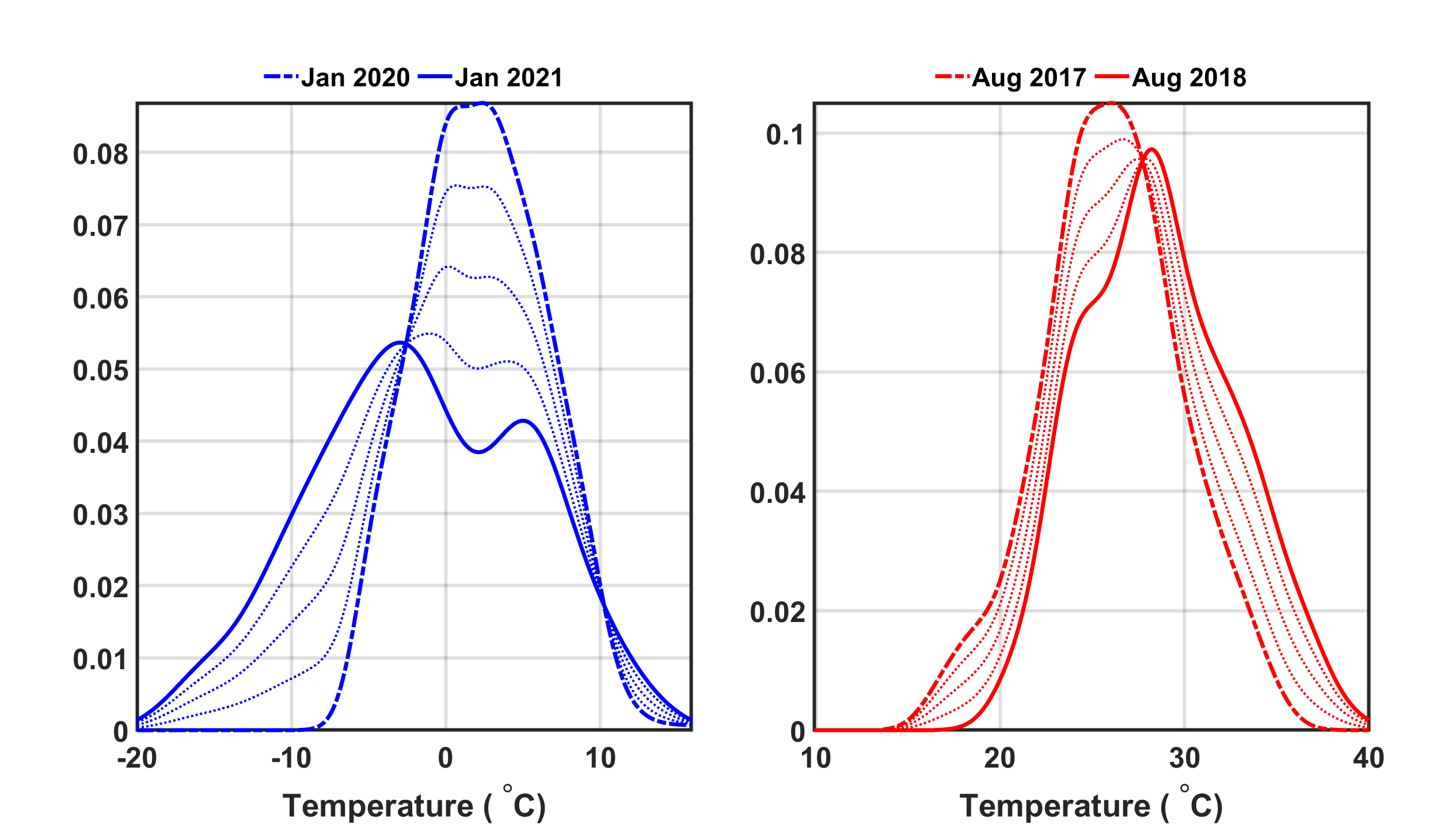}
\includegraphics[height=0.31\textwidth, width=0.45\textwidth]{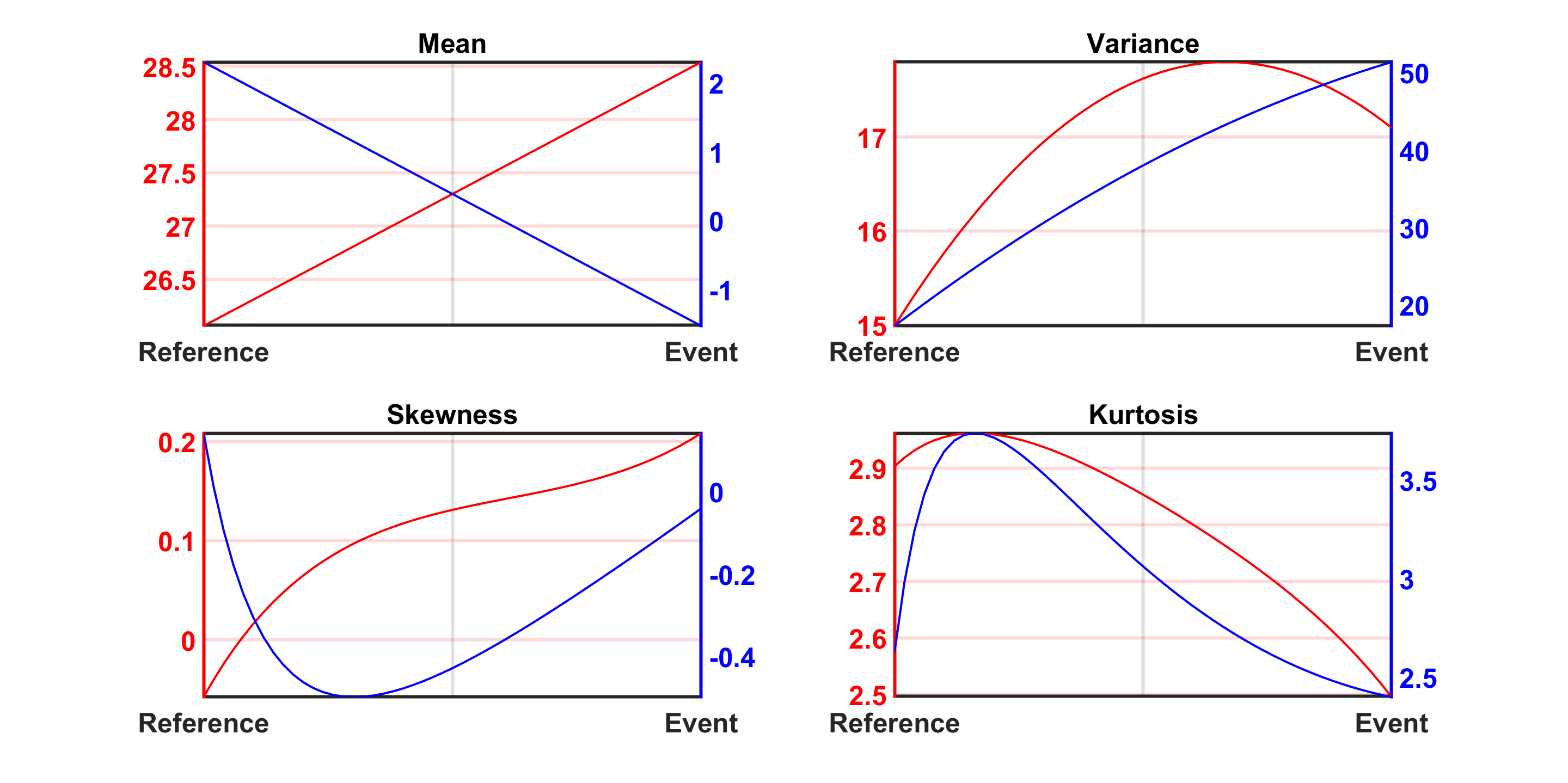}
\caption{Temperature PDFs for cold waves and heat waves (left), showing the transition from the reference to event periods, and the evolution of descriptive statistics (right).}
\label{Fig:CLR_Func_shock0}
\end{center}
\end{figure}

\indent Our empirical analysis aims to provide practitioners with insights into the potential increase in future electricity demand when similar extreme temperature events occur. By understanding extreme temperature events as potential perturbations in the distributional predictor \( X_t \), we can estimate their impact on electricity demand. We henceforth refer to the function \( \zz_{\ee}(s) \), corresponding to the black line in the left (resp.\ right) panel of Figure \ref{Fig:Func_shock2}, as the observed cold wave (resp.\ heat wave). This function is computed by subtracting \( \ZZ_{\nn}(s) \) from \( \ZZ_{\ee}(s) \), where \( \ZZ_{\nn}(s) \) denotes the reference (normal) temperature PDF in January 2020 (resp.\ August 2017), and \( \ZZ_{\ee}(s) \) denotes the extreme temperature PDF in January 2021 for the cold wave (resp.\ August 2018 for the heat wave). Consequently, \( \ZZ_{\ee}(s)  =  \ZZ_{\nn}(s) + \zz_{\ee}(s) \). 

\indent Due to the flexibility of our model in the choice of potential perturbations, we may also consider fractions of these observed heat and cold waves and their effects on electricity demand; specifically, we let 
\begin{equation} \label{eqfraction}
\zz_a(s) = a \times \zz_{\ee}(s), \quad a = \frac{1}{M}, \frac{2}{M}, \ldots, 1, 
\end{equation}
for some large positive integer \( M \). Note that these fractions of observed heat or cold waves satisfy (i) \( |\zz_a(s)| \leq |\zz_{\ee}(s)| \) and (ii) \( \int_{-20}^{40} \zz_a(s) \, ds = 0 \) for all considered values $a$. Thus, adding these fractions of heat or cold waves to the reference PDF \( \ZZ_{\nn}(s) \) results in a new PDF, hereafter denoted as \( \ZZ_a(s)\) (i.e., \(\ZZ_a(s) = \ZZ_{\nn}(s) + \zz_a(s)) \). Since the application of \( \zz_a(s) \) induces a globally more moderate shock compared to that of \( \zz_{\ee}(s) \), \( \ZZ_a(s) \) represents less extreme temperature conditions than the extreme temperature PDF \( \ZZ_{\ee}(s) \) observed in January 2021 for the cold wave or August 2018 for the heat wave. Thus, \( \zz_a(s) \) can naturally be interpreted as a possible, but less extreme, heat or cold wave. Observe also that, as $a$ grows, $\ZZ_a(s)$ constructed from $\zz_a(s)$ moves away from the normal temperature PDF ($\ZZ_{\nn}(s)$) and closer to the extreme temperature PDF ($\ZZ_{\ee}(s)$) by a constant functional increment $(1/M) \times \zz_{\ee}(s)$ at each step; the left panel of Figure \ref{Fig:CLR_Func_shock0} illustrates $\ZZ_a$ for a few selected values of $a$.

\begin{figure}[t]
\begin{center}
\includegraphics[height=0.33\textwidth, width=0.42\textwidth]{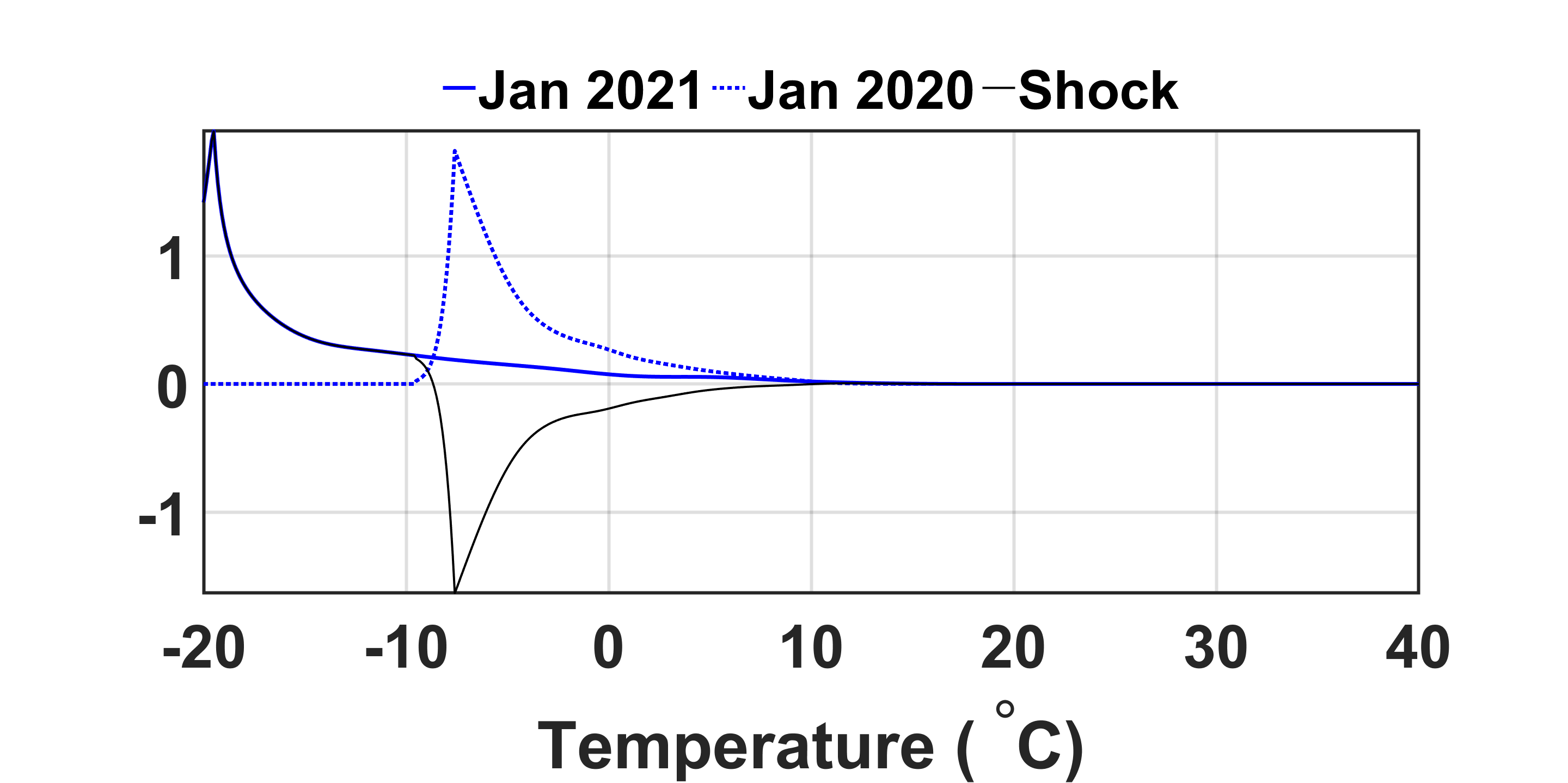}
\includegraphics[height=0.33\textwidth, width=0.42\textwidth]{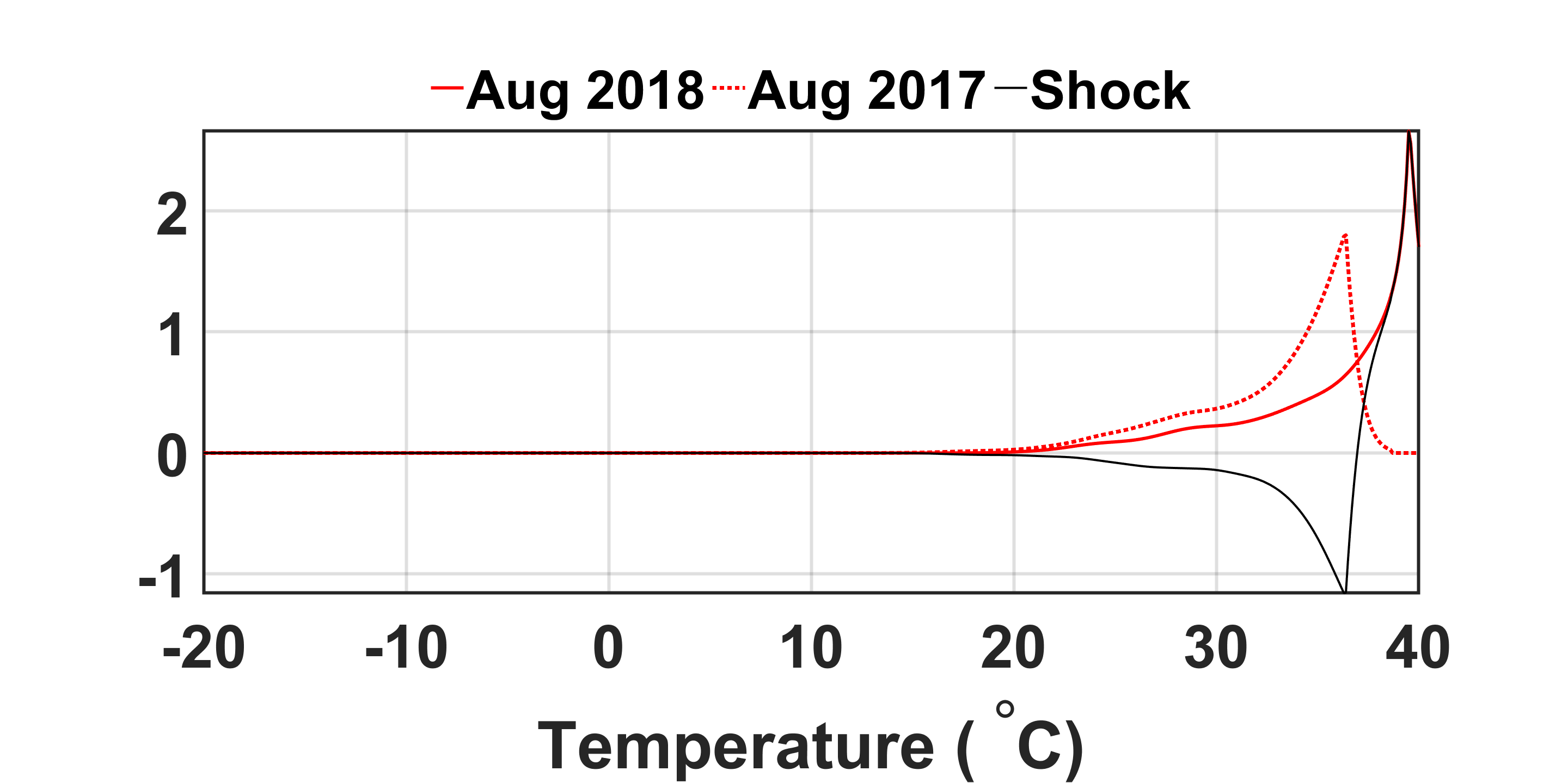} \\
\includegraphics[height=0.33\textwidth, width=0.42\textwidth]{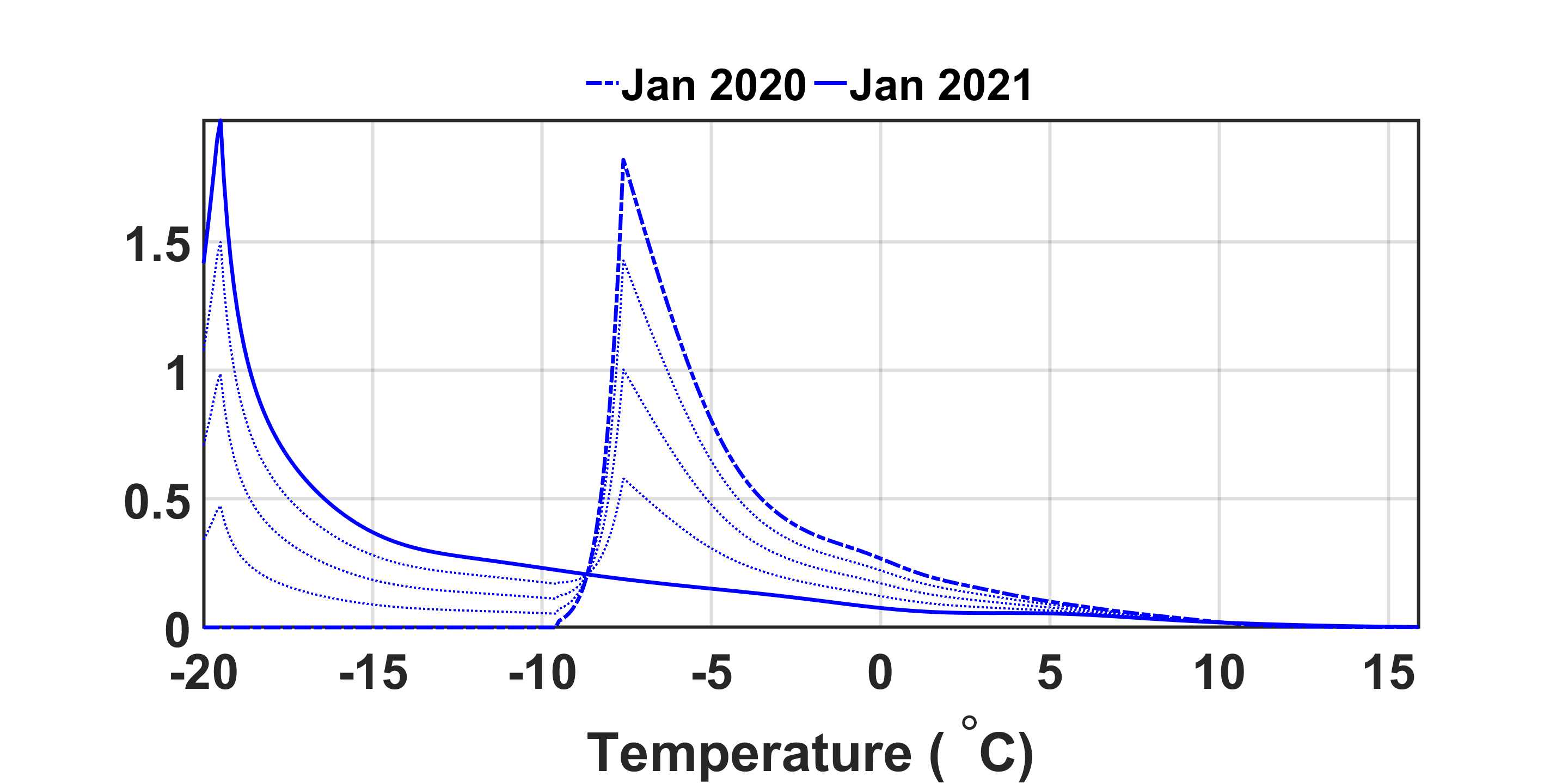}
\includegraphics[height=0.33\textwidth, width=0.42\textwidth]{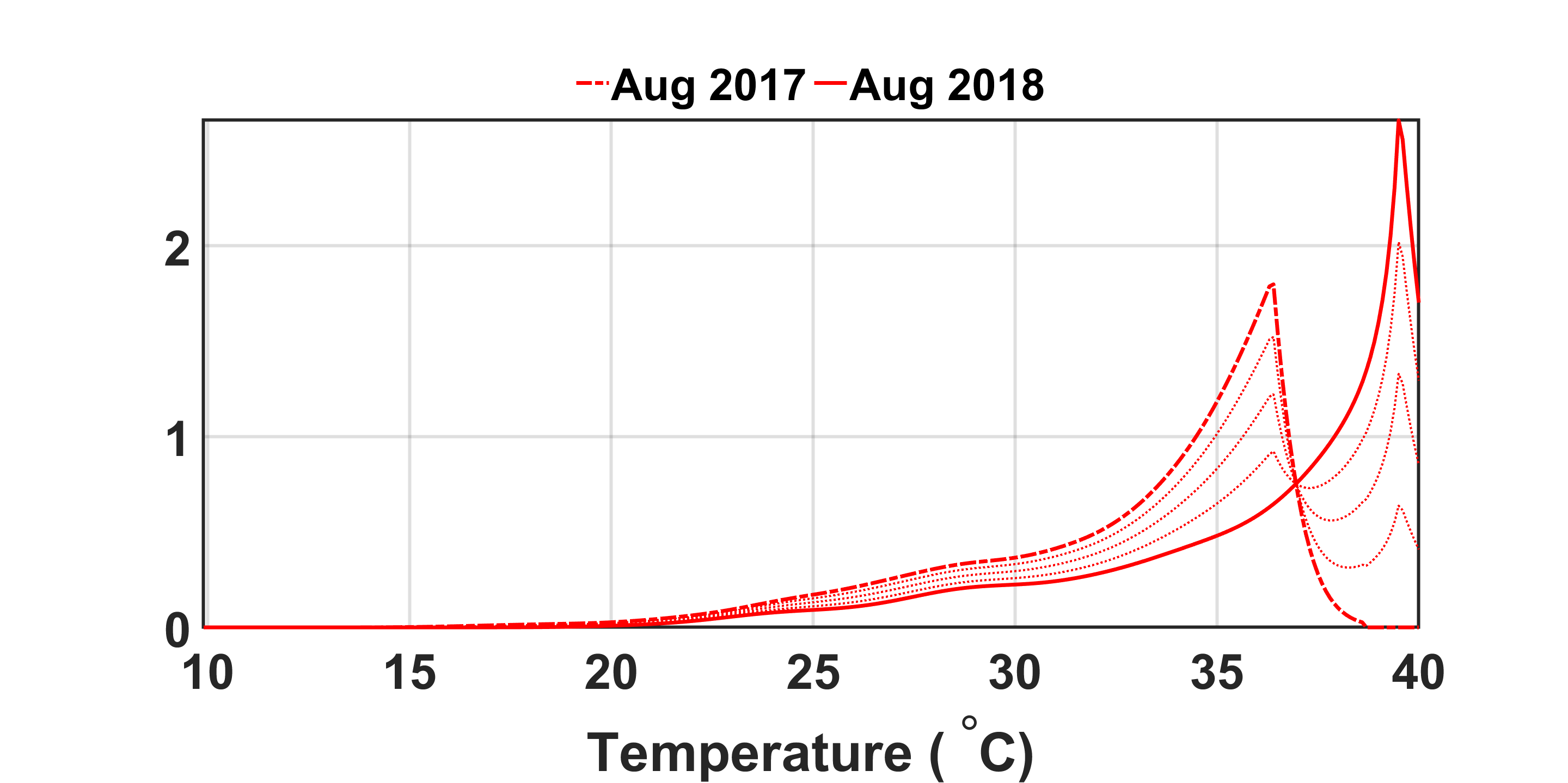}
\caption{Estimated LRHR and LHR functions for cold waves (top left) and heat waves (top right). The transition from the reference to event periods by functional LRHR and LHR shocks is depicted in the bottom panels.} 
\label{Fig:HtoD_Shocks} 
\end{center}
\end{figure}

\indent Similarly, Figure \ref{Fig:HtoD_Shocks} illustrates the corresponding year-over-year changes in monthly temperature LRHR and LHR for these cold and heat wave events, respectively. Specifically, in January 2021, the LRHR gradually increased as temperatures decreased, peaking at -19.5°C, which represents the highest conditional probability of occurrence at this temperature. This indicates that, given temperatures had already dropped below -19.5°C, the likelihood of further decreases to even lower temperatures was greatest. This peak highlights -19.5°C as a critical threshold during the cold wave events. In contrast, January 2020 exhibited a peak LRHR at -7.6°C, reflecting a focus on milder cold conditions and a rapid decline in LRHR at more extreme lows. Similarly, in August 2018, the LHR peaked at 39.5°C, indicating the dominance of intense heatwave conditions. In comparison, August 2017 exhibited a peak at 36.4°C, indicating a milder concentration of high-temperature events. It is worth noting that the exponentially increasing pattern of the monthly temperature LHR in August 2018 (LRHR in January 2021) indicates that the probability of the temperature staying around 39.5°C (-19.5°C) is higher than for any temperature below (above) this value. This would demonstrate the cumulative temperature effect on electricity demand during peak times to some extent (see Section \ref{sec_more}). The bottom left (resp.\ right) panel of the figure illustrates the transition from the normal temperature LHR (resp.\ LRHR) to the extreme temperature LHR (resp.\ LRHR). 

\begin{figure}[t]
\begin{center}
\includegraphics[height=0.3\textwidth, width=0.66\textwidth]{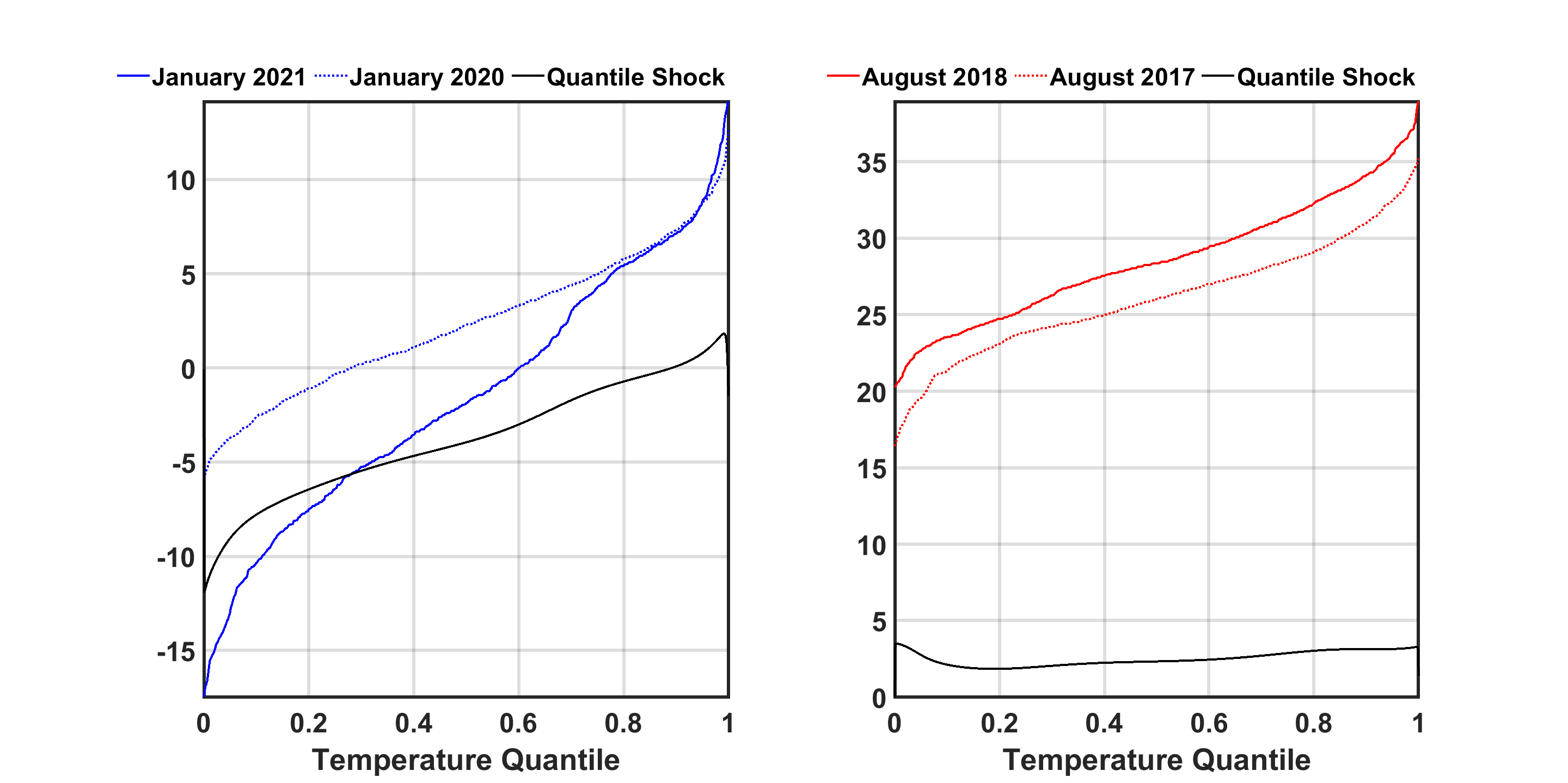}
\includegraphics[height=0.3\textwidth, width=0.33\textwidth]{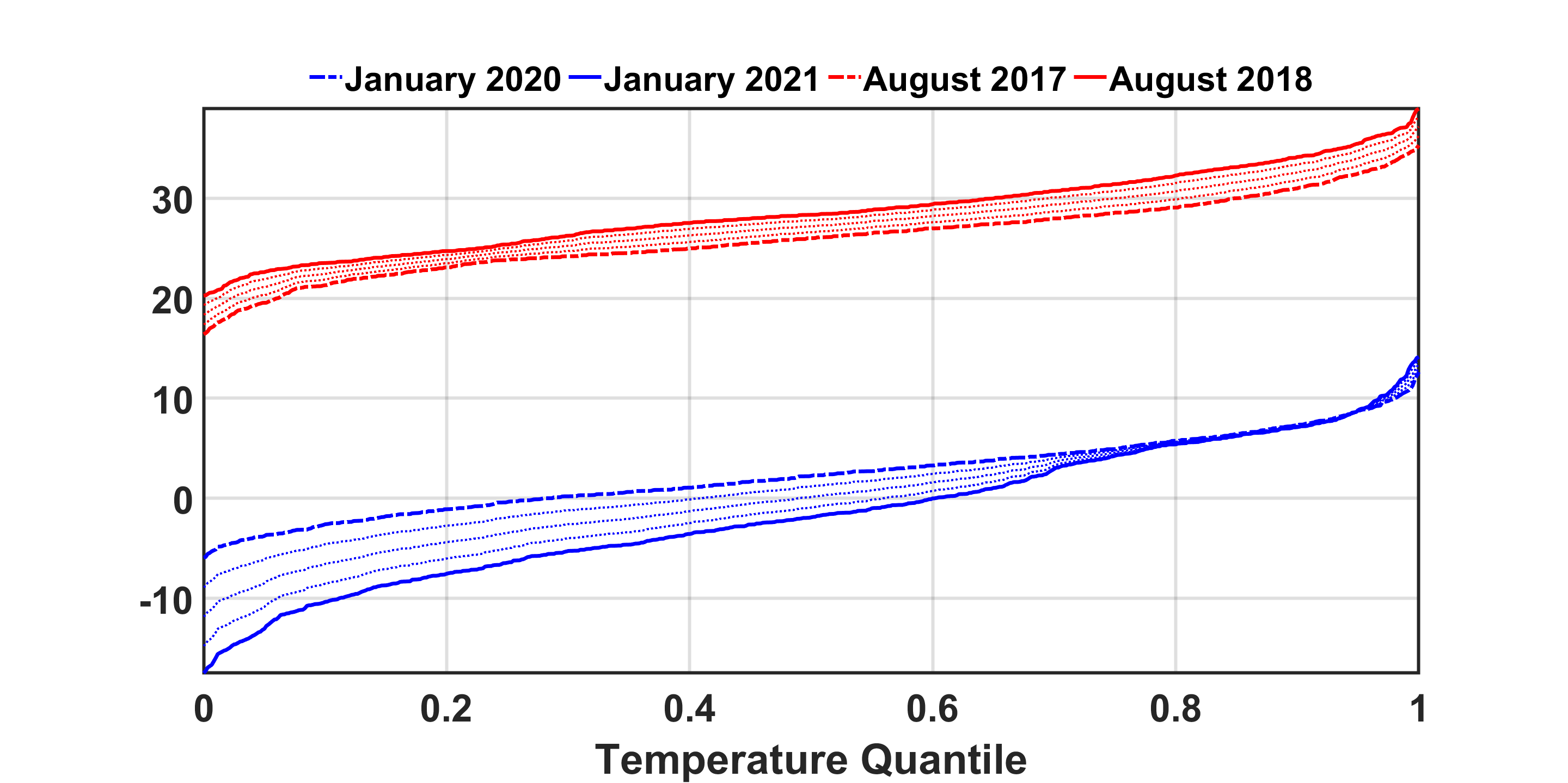}
\caption{Estimated quantile functions for cold waves (left) and heat waves (middle), and the transition from the reference to event periods by functional quantile shocks (right).} 
\label{Fig:Quantile_Shock}
\end{center}
\end{figure}

Given that the quantile function corresponding to the temperature distribution can serve as a distributional summary, it may be of interest to practitioners to examine changes in quantile functions during the considered extreme temperature events. The left panel of Figure \ref{Fig:Quantile_Shock} shows the quantile functions corresponding to the normal and extreme temperature distributions and also illustrates the cold wave event, expressed as a quantile shock. As shown, the impact is most pronounced in the lower quantiles, particularly those associated with the coldest temperatures. This reflects that the January 2021 cold wave caused a significant downward shift in the quantile function compared to January 2020, with the most substantial changes at the lowest quantiles. In the middle panel, the quantile shocks associated with the August 2018 heat wave exhibit a relatively uniform increase across the quantiles, with a more pronounced effect at both extremes. This implies that the heat wave not only elevated overall temperature levels but also disproportionately intensified occurrences at both the lower and upper quantiles. The fractions of observed cold and heat waves, expressed as quantile shocks, can be constructed as in \eqref{eqfraction}. These quantile shocks are used to illustrate the transition from the normal temperature quantile function to the extreme one in each scenario, as shown in the right panel of the figure. 

\bigskip
\subsection{Empirical Results for Density-to-Demand Model}\label{DtoD_Rlt}
\noindent In this section, we analyze the impact of extreme temperature events on residential electricity demand using the PDF-based DSR model, hereafter referred to as the density-to-demand model. We first apply this model to estimate point-wise temperature sensitivity through the benchmark response function. Subsequently, we analyze the effects of observed heat and cold wave fractions on electricity demand using the CLR predictor. Note that we generate the relevant functional shocks $\zeta_a$, corresponding to \( \zz_a(s) \), in the CLR domain. This can be implemented by applying the CLR transformation to both the reference PDF \( \ZZ_{\nn}(s) \) and \( \ZZ_a(s) \), and then computing the subtraction of the former from the latter in each of the heat and cold wave scenarios.\footnote{If $\mathrm{X}_{\ee}$ (resp.\ $\mathrm{X}_{a})$ is the CLR  corresponding to $\ZZ_{\ee}$ (resp.\ $\ZZ_{a})$, then $\zeta_a=\mathrm{X}_{\ee} - \mathrm{X}_{a}$.} The effect of $\zeta_a$ on electricity demand is then given by $f(\zeta_a)$. 

\begin{figure}[t]
\begin{center}
\includegraphics[height=0.35\textwidth, width=0.41\textwidth]{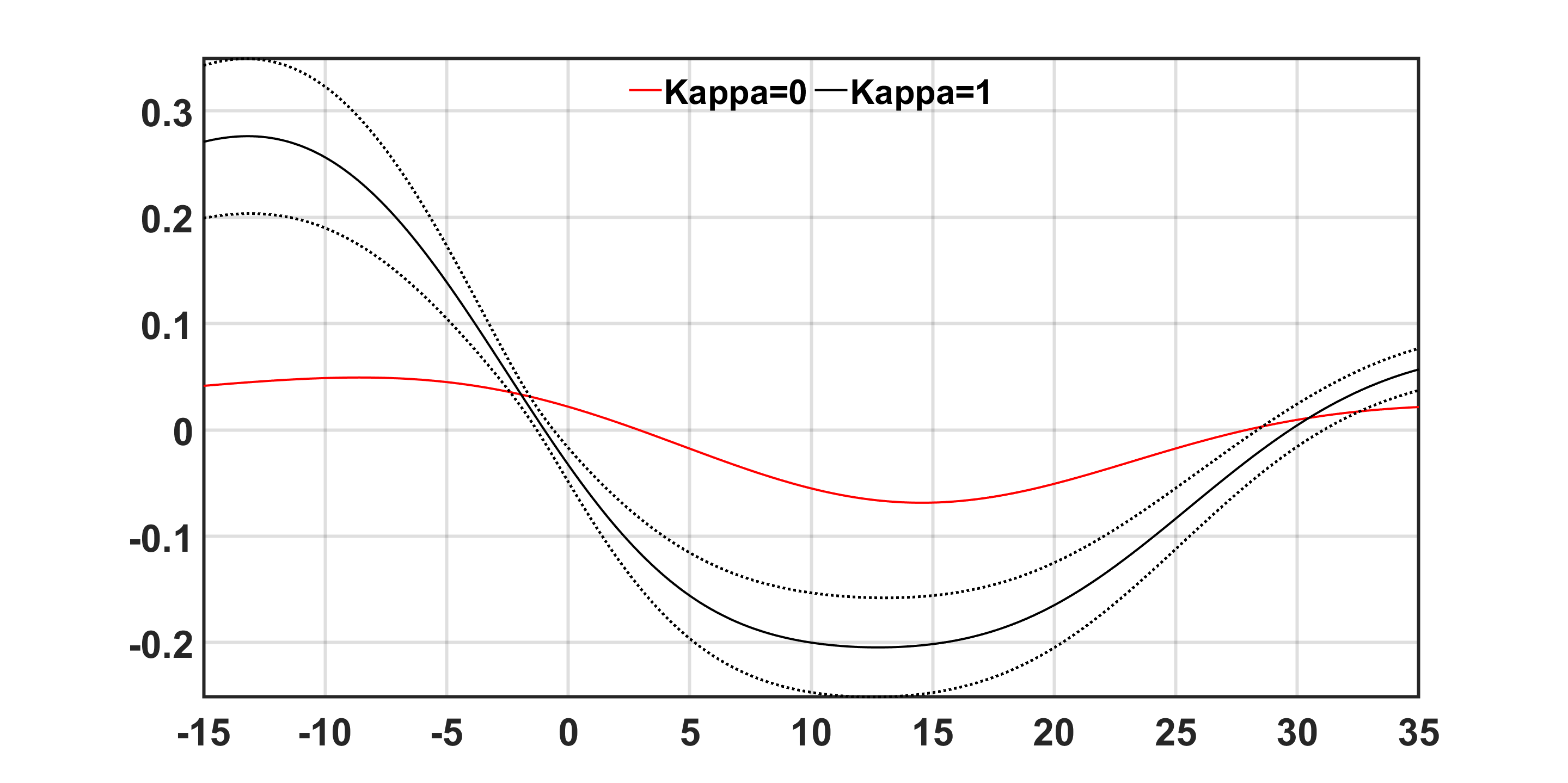}
\includegraphics[height=0.35\textwidth, width=0.58\textwidth]{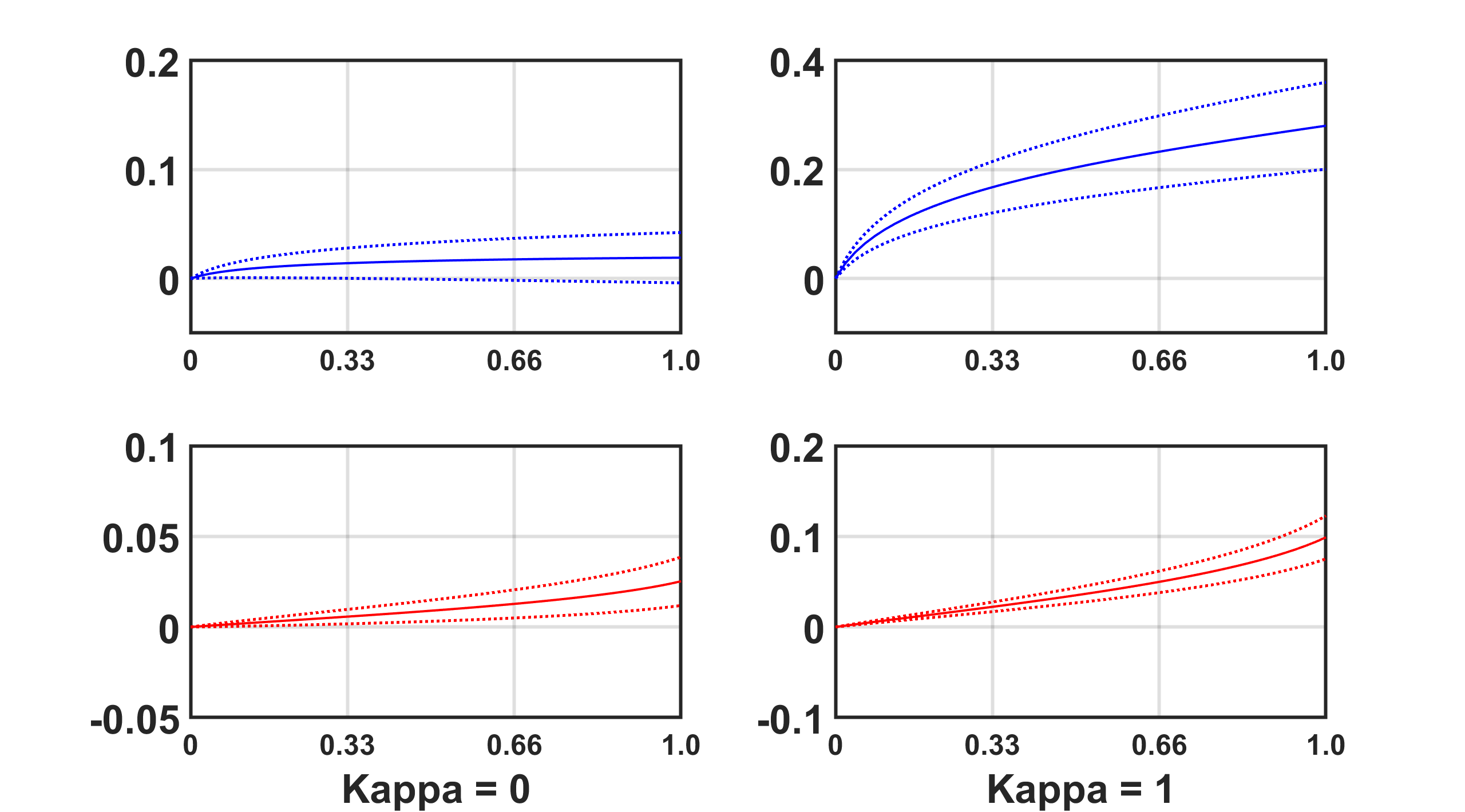}
\caption{Estimated PDF-based benchmark response function with 95\% confidence intervals (left), and PDF-based temperature sensitivity for varying cold wave (top middle and top right) and heat wave (bottom middle and bottom right) shock intensities under $\kappa = 0$ and $\kappa = 1$.}
\label{Fig:Empirical_rlts}
\end{center}
\end{figure}

\indent In the left panel of Figure \ref{Fig:Empirical_rlts}, we present our benchmark response function on the range $s \in [-15,35]$ and its 95\% confidence interval. Consistent with the findings of \cite{chang2016new}, the estimated temperature response function exhibits a nonlinear U-shaped pattern across the temperature domain. However, the estimates with $\kappa=1$ reveal a more pronounced degree of nonlinearity. Note also that, unlike existing approaches, our approach explicitly accounts for the interdependence of temperature changes, ensuring that adjustments at one level are accompanied by compensatory changes in other temperatures to maintain the unit integral requirement of a PDF.\footnote{The function $f(\zeta_s)$, where $\zeta_s$ is defined near $s$, represents the effect of an additional perturbation applied to the distributional predictor $X_t$, as indicated by $f(X_t+\zeta_s) - f(X_t)$. If $X_t$ is the PDF itself, $X_t+\zeta_s$ may not be a valid PDF, as it could violate the unit integral requirement. However, this issue does not arise if $X_t$ is the CLR transformation of a PDF and $\zeta_s$ is a shock expressed in the CLR.}


\indent The right two panels of Figure \ref{Fig:Empirical_rlts} present the estimated response of electricity demand on the $y$-axis with respect to $a$, denoting the fraction parameter defined in \eqref{eqfraction}, of $\ZZ_a(s)$, depicted on the $x$-axis. As earlier discussed, $\ZZ_a(s)$ approaches $\ZZ_{\ee}(s)$ with a constant functional increment as $a$ increases. The solid lines show the estimates, $\hat{f}(\zeta_a)$, which can be interpreted as the estimated demand responses to progressively but constantly intensifying heat or cold wave shocks, as described in Section \ref{Sec_Est_Shocks}. The dotted lines indicate the 95\% confidence intervals for $f(\zeta_a)$ based on the asymptotic normality result given in \eqref{eqlocal}. 

\indent The empirical results from the estimation with $\kappa = 1$, which is designed to account for potential measurement errors, indicate an increase in electricity demand response as the considered shock gets closer to the observed cold or heat wave. For comparison, we also report estimation results without accounting for potential endogeneity, with the $y$-axis scaled to half of that in the case of $\kappa=1$ to enhance visualization. Similar to the benchmark response function, the empirical results for observed heat and cold wave fractions with $\kappa=0$ exhibit a similar upward trend. However, the magnitude is substantially lower compared to the case with $\kappa=1$. Since the estimation results for $\kappa=0$ do not account for endogeneity, we focus on interpreting the estimation results with $\kappa=1$ as our main empirical findings.  

\indent Note that the right panel of Figure \ref{Fig:CLR_Func_shock0} provides statistical insights into how the considered cold and heat wave events change temperature distributions. As the considered cold wave shock approaches the observed severe cold wave, the mean temperature experiences a linear decline from 2.31°C to -1.53°C, with a substantial increase in variance ($17.30 \rightarrow 51.37$), indicating greater temperature dispersion. Concurrently, skewness decreases ($0.14 \rightarrow -0.04$), and kurtosis declines ($2.62 \rightarrow 2.40$), signifying a more symmetrical and less peaked distribution. In contrast, as the considered heat wave shock approaches the observed severe heat wave, the mean temperature rises linearly from 26.02°C to 28.54°C, with a modest increase in variance ($15.0 \rightarrow 17.11$), rising skewness ($-0.06 \rightarrow 0.20$), and declining kurtosis ($2.90 \rightarrow 2.50$), reflecting a more right-skewed but less peaked distribution. As illustrated in the right panel of Figure \ref{Fig:Empirical_rlts}, these distributional changes, characterized by greater variance shifts per unit change in mean temperature, reveal that cold wave shocks have a significantly greater impact on electricity demand, with a 28.06\% increase, compared to a 9.89\% increase observed during heat wave shocks.

\indent The observation that cold wave shocks lead to a greater increase (7.31\%) in electricity demand per unit change in mean temperature, following the transition from $\ZZ_{\nn}$ to $\ZZ_{\ee}$, compared to heat wave shocks (3.93\%) highlights that electricity demand is more sensitive to the extreme and intense temperature conditions associated with cold waves. This implies that demand response is influenced not only by temperature extremes but also by the specific nature of these extremes. This finding departs from existing literature, which predominantly  focused on the demand response to changes in temperature levels alone, and also highlights the importance of understanding how shifts in temperature distributions impact electricity demand. Such insights emphasize the need for energy planning that accounts for the varying effects of temperature distribution changes, particularly in the context of increasingly severe cold wave events.

\indent Notably, as the cold waves intensify, electricity demand rises at a diminishing rate as incremental heating needs taper off. More specifically, the demand response values for the cold wave event at fraction parameters $a$ of 0.33, 0.66, and 1.0 are estimated at 0.1677, 0.2328, and 0.2806, respectively. Given that electricity demand is expressed in natural logarithms, these values indicate that a functional shock causing normal temperatures to approach extreme levels by 33\% corresponds to a 16.77\% year-over-year increase in monthly electricity demand. An additional 33\% progression toward extreme temperatures results in a further 6.51\% increase, followed by an additional 4.77\% increase with a complete transition to extreme conditions. Collectively, the full transition from normal temperatures in January 2020 to extreme temperatures in January 2021 corresponds to a total 28.06\% year-over-year increase in monthly electricity demand.

\indent Conversely, electricity demand is estimated to increase at an accelerating rate as the heat waves intensify, driven by heightened cooling requirements. More specifically, the demand response values for the heat wave events at fraction parameters $a$ of 0.33, 0.66, and 1.0 are 0.0222, 0.0498, and 0.0989, respectively. Using the same approach, these results indicate that a functional shock causing normal temperatures to approach extreme levels by 33\% corresponds to a 2.22\% year-over-year increase in monthly electricity demand. An additional 33\% progression toward extreme temperatures results in a 2.76\% year-over-year increase, followed by an additional 4.91\% increase with the complete transition to extreme conditions. Collectively, the full transition from normal temperatures in August 2017 to extreme temperatures in August 2018 corresponds to a total 9.89\% year-over-year rise in monthly electricity demand. This contrasting behavior can be attributed to the extensive reliance on air conditioning during heat waves and the substitution effect between natural gas and electricity for heating purposes during cold waves. 

\bigskip
\subsection{Empirical Results for Hazard-to-Demand Model}\label{DtoD_Rlt2}
\begin{figure}[t]
\begin{center}
\includegraphics[height=0.2\textwidth, width=0.41\textwidth]{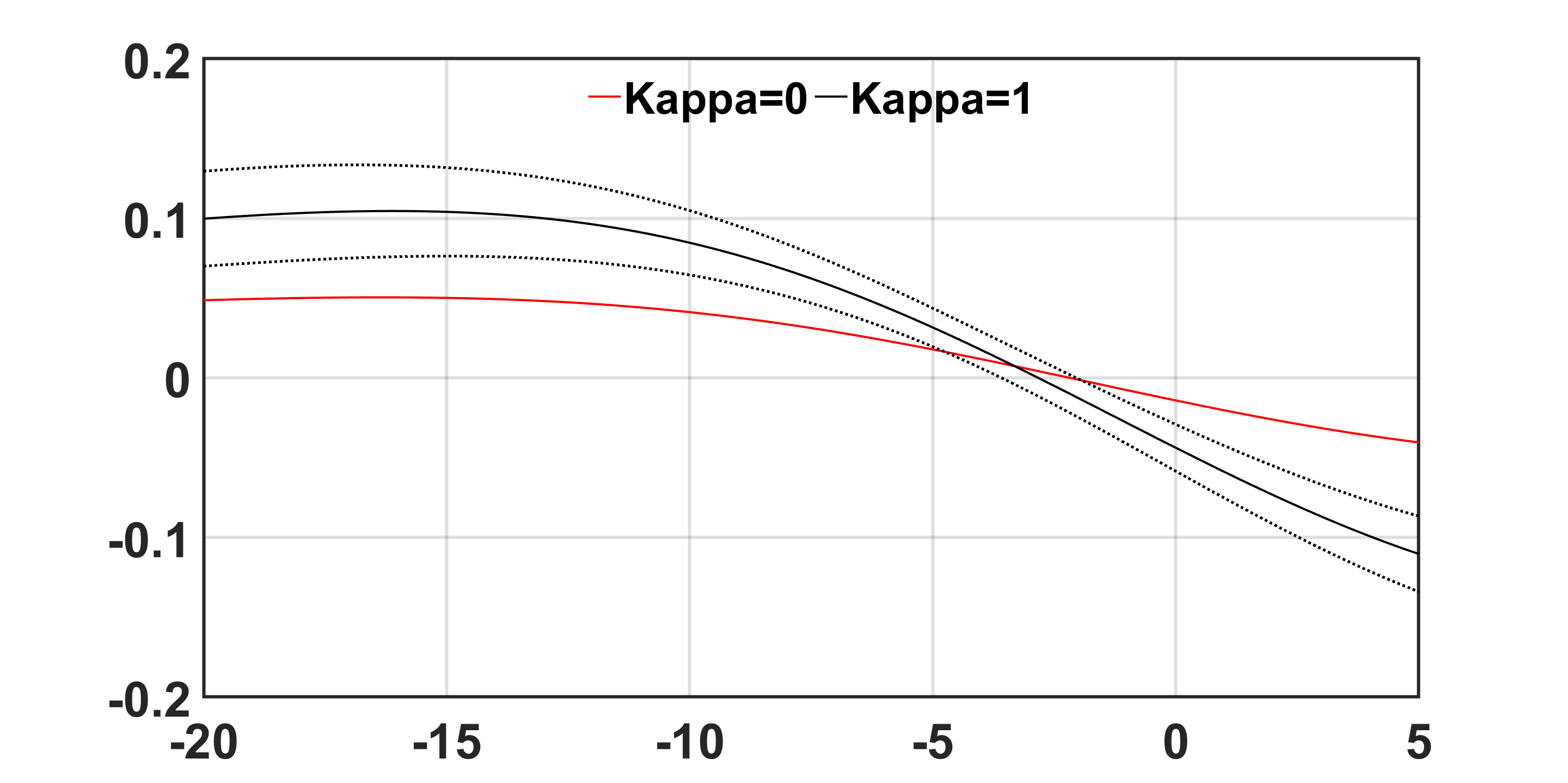}
\includegraphics[height=0.2\textwidth, width=0.58\textwidth]{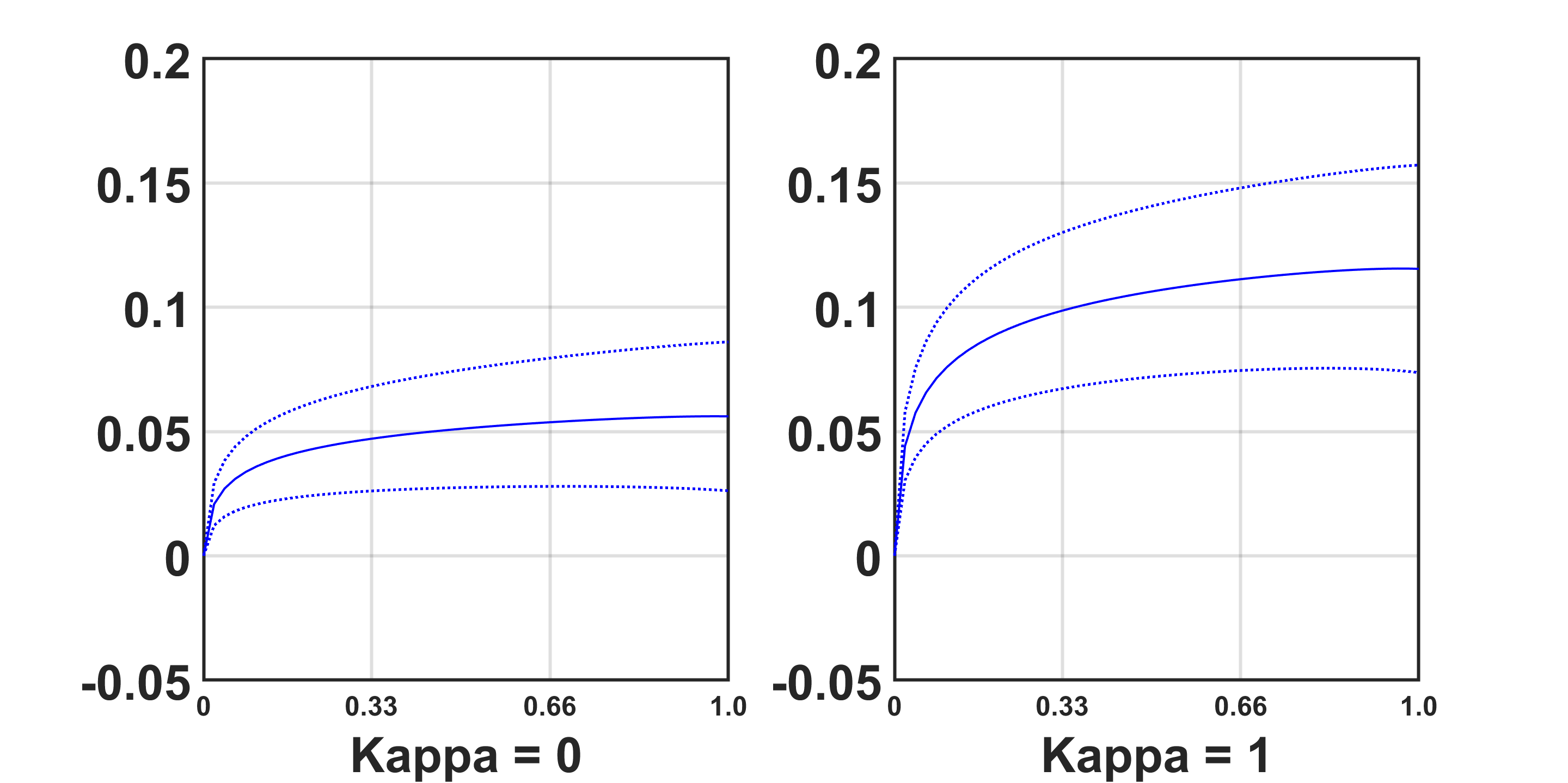}
\includegraphics[height=0.2\textwidth, width=0.41\textwidth]{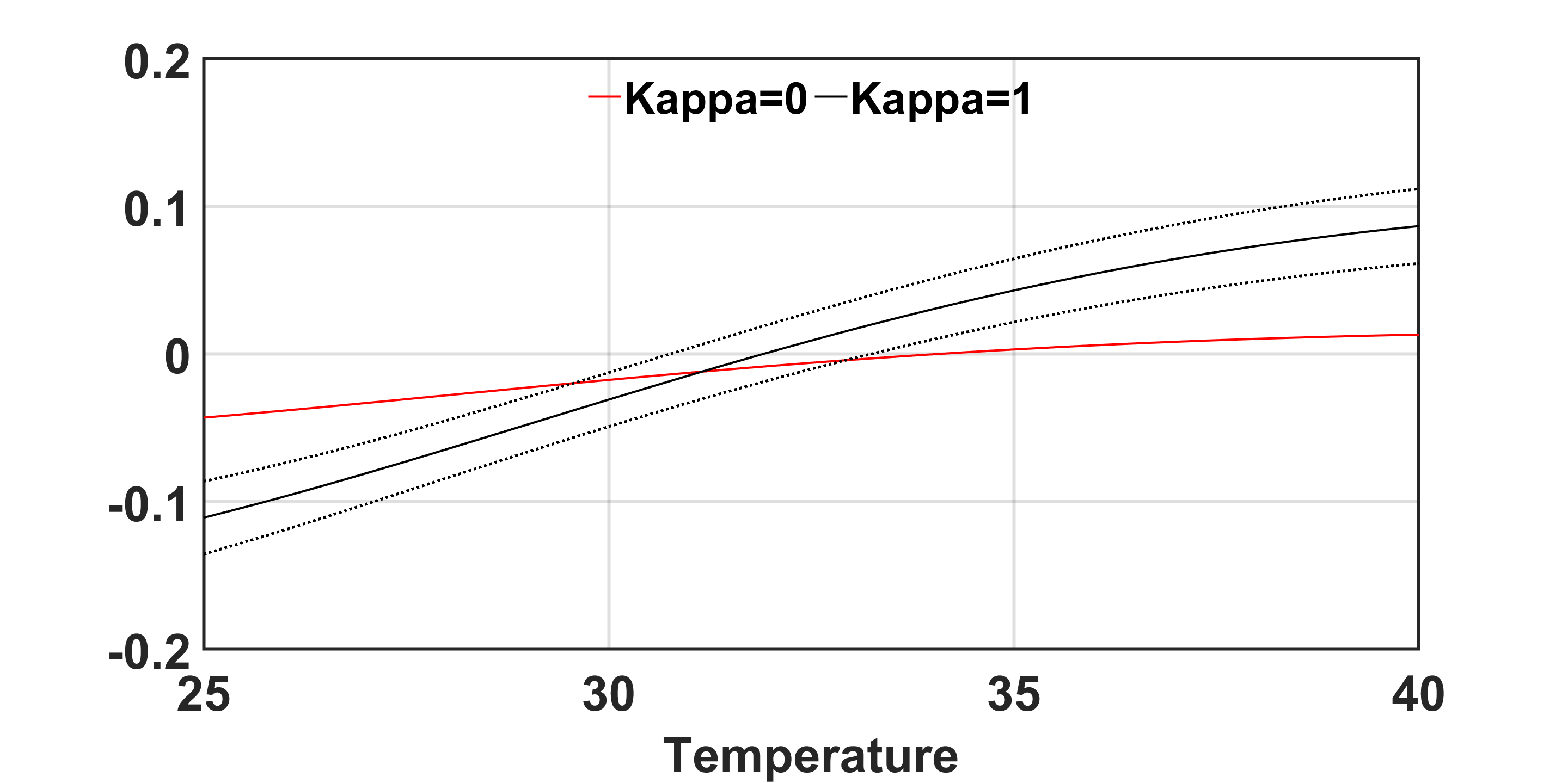}
\includegraphics[height=0.2\textwidth, width=0.58\textwidth]{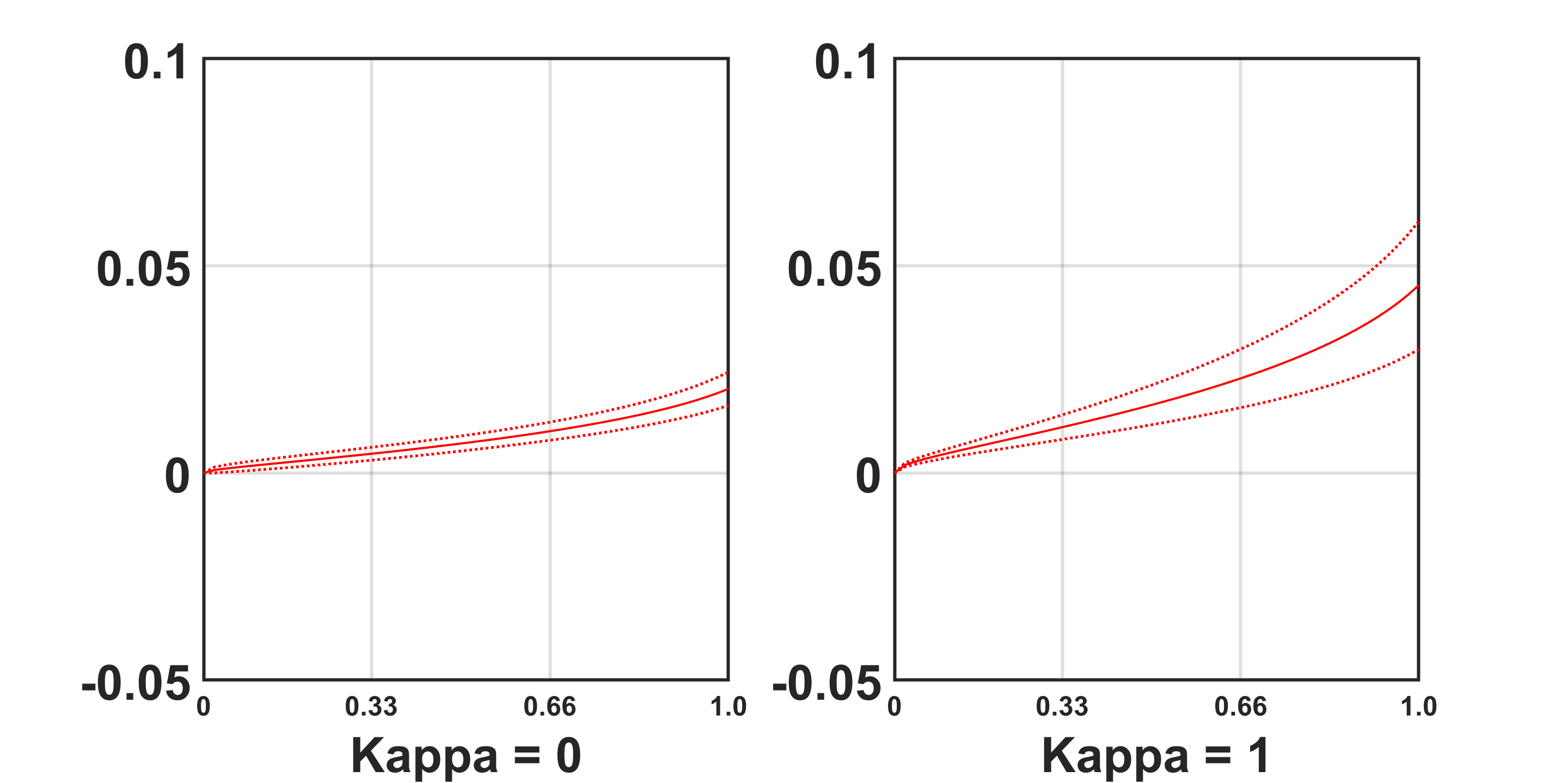}
\caption{Estimated LHR-based benchmark response function with 95\% confidence intervals (left), and LHR-based temperature sensitivity for varying cold wave (top middle and top right) and heat wave (bottom middle and bottom right) shock intensities under $\kappa = 0$ and $\kappa = 1$.}
\label{Fig:HtoD_rlts}
\end{center}
\end{figure}

\noindent In this section, we analyze the impact of extreme temperature events on residential electricity demand using the DSR model with the LHR (resp.\ LRHR) predictor for both hypothetical and historical heat (resp.\ cold) wave scenarios, collectively referred to as the hazard-to-demand model. Figure \ref{Fig:HtoD_rlts} illustrates the benchmark response function $\widetilde{\psi}(s)$, focusing on the extreme temperature ranges of $[-20, 5]$ and $[25, 40]$, and the corresponding estimated electricity demand response as a function of parameter $a$ within the hazard-to-demand model. While the estimation results from the hazard-to-demand model are broadly consistent with those from the density-to-demand model, they provide distinct interpretative insights into how variations in the conditional probability of extreme temperature events distinctly influence electricity demand under cold and heat wave scenarios.

\indent While both models capture increased temperature sensitivity, the hazard-to-demand model produces higher response function values in the extreme heat range but lower values in the extreme cold range compared to the density-to-demand model. Notably, it provides a more reliable representation of the response function at the lower temperature boundary, offering a structured depiction of cold extremes. Specifically, electricity demand rises as temperatures decline within the cold temperature range. However, at extremely low temperatures, demand exhibits diminishing sensitivity to further declines, reflecting saturation effects driven by persistent heating requirements and substitution effects with natural gas. In contrast, in the hot temperature range, rising temperatures correspond to an increased conditional probability of extreme heat events, leading to a steady rise in electricity demand, particularly at higher temperatures. This increased sensitivity is primarily driven by intensive cooling needs, highlighting the asymmetric response of electricity demand to temperature extremes. Further discussion on these distinctions is provided in the following section. 

\indent Although with a smaller overall magnitude, the hazard-to-demand model aligns with the density-to-demand model for historical extreme temperature events. The cold wave shocks result in a greater increase in electricity demand per 1°C change in mean temperature (3.01\%) following the transition from $\ZZ_{\nn}$ to $\ZZ_{\ee}$, compared to the heat wave shocks (1.80\%). Moreover, as the cold waves intensify, the rate of increase in electricity demand diminishes, reflecting tapering incremental heating needs. Specifically, the LRHR-based demand response values for the cold wave events at fraction parameters $a$ of 0.33, 0.66, and 1.0 are estimated at 0.0987, 0.1113, and 0.1155, respectively. These estimates indicate that a functional shock causing normal temperatures to shift toward extreme cold levels by 33\% corresponds to a 9.87\% year-over-year increase in monthly electricity demand. An additional 33\% progression toward extreme cold conditions results in a further 1.26\% increase, followed by an additional 0.42\% increase with a complete transition to extreme cold conditions.

\indent For the heat wave events, the LHR-based demand response values at fraction parameters $a$ of 0.33, 0.66, and 1.0 are estimated at 0.0111, 0.0228, and 0.0453, respectively. These results indicate that a functional shock causing normal temperatures to shift toward extreme heat levels by 33\% corresponds to a 1.11\% year-over-year increase in monthly electricity demand. An additional 33\% progression toward extreme heat conditions results in a further 1.17\% increase, followed by an additional 2.25\% increase with a complete transition to extreme heat conditions. Both the density-to-demand and hazard-to-demand models, despite differences in their estimated magnitudes, consistently reveal distinct demand response patterns for the heat wave and cold wave events. The cold wave shocks result in more significant demand increases overall while the heat wave shocks exhibit a stronger incremental effect as temperatures approach extreme levels. 

\indent Lastly, similar to the density-to-demand model, the empirical results from the hazard-to-demand model for $\kappa=0$ exhibit an upward trend but with a substantially lower magnitude compared to the case with $\kappa=1$. As the estimation results for $\kappa=0$ do not account for endogeneity, they are not considered for interpretation.

\bigskip
\subsection{Some Discussion on the Distributional Predictors} \label{sec_more}

\noindent Our empirical findings for extreme temperature events provide more practical insights than those derived from existing functional approaches. A key limitation of conventional methods is their tendency to yield inaccurate estimates at the boundaries, particularly in data-sparse regions near critical temperature thresholds such as -20°C and 40°C. This boundary issue can lead to misinterpretations of demand responses at extreme temperatures, potentially resulting in significant misjudgments in energy policy and management.

\indent By leveraging both representative heat and cold wave scenarios, as well as demand responses at specific temperature levels, our approach provides broader distributional insights into electricity demand dynamics under extreme temperature events. Specifically, we incorporate higher-order distributional changes associated with extreme temperatures, illustrating how such conditions result in significantly greater increases in electricity demand compared to normal scenarios--an effect that has been largely overlooked in previous studies. Additionally, we integrate the LHR and LRHR predictors, which are expected to more effectively capture electricity demand dynamics under extreme temperature conditions, into the hazard-to-demand model, a framework not previously explored in the literature. Furthermore, our estimation results remain robust to endogeneity concerns arising from measurement errors in the distributional predictors.


The estimation results from both models, as presented in Figures \ref{Fig:Empirical_rlts} and \ref{Fig:HtoD_rlts}, indicate that the benchmark response function from the hazard-to-demand model more effectively captures electricity demand dynamics under extreme temperature conditions than that of the density-to-demand model. In the extreme cold range, the benchmark response function from the hazard-to-demand model (top left panel of Figure \ref{Fig:HtoD_rlts}) remains non-decreasing, except below $-16.2$°C, a threshold not observed in historical cold wave events. In contrast, the benchmark response function from the density-to-demand model (left panel of Figure \ref{Fig:Empirical_rlts}) begins to decline at $-13.3$°C, suggesting an implausible reduction in electricity demand under extreme cold temperatures. Similarly, in the extreme heat range, the benchmark response function from the hazard-to-demand model suggests a steeper increase in demand, surpassing the growth rate implied by that of the density-to-demand model. 

While benchmark response functions from both models consistently show that cold waves result in greater demand increases than heat waves, that of the density-to-demand model exhibits excessive sensitivity, predicting disproportionately large demand fluctuations in response to imperceptibly small temperature variations (e.g., 0.1°C) under extreme cold conditions. In contrast, the benchmark response function from the hazard-to-demand model under extreme cold scenarios demonstrates more moderate sensitivity but still exhibits slightly larger responses compared to extreme heat scenarios. In this context, the hazard-to-demand model aligns more closely with empirical observations from historical heat and cold wave events.


\indent In the reported estimation results for the functional shocks generated from the observed cold and heat waves (the right panels of Figures \ref{Fig:Empirical_rlts} and \ref{Fig:HtoD_rlts}), it is important to note that both the density-to-demand and hazard-to-demand models capture the incremental effects of extreme heat on electricity demand, whereas the benchmark response function \( \widetilde{\psi}(s) \) from either of the models fails to capture such effects as temperature approaches extreme heat. This suggests that distributional characteristics associated with heat wave events, such as a more right-skewed distributional change with significantly reduced kurtosis, further amplify electricity demand in a nonlinear manner. In contrast, incorporating distributional characteristics of historical cold wave events, such as significantly increased variance, into the benchmark response function does not qualitatively affect its overall shape.


Based on the derived implications above, the hazard-to-demand model appears to provide an empirically more sensible framework for analyzing the impact of extreme temperature events on electricity demand. This may be because, unlike the PDF, which considers the entire temperature distribution and may consequently dilute the influence of rare but critical extreme events that occur as tail events, the LHR predictor assigns substantial importance to tail events, as is evident from its construction. This allows for a more targeted analysis, focusing on the likelihood of temperatures persisting within an extreme range given that they have already exceeded a critical threshold (e.g., -15°C or 35°C). By emphasizing tail events, the LHR approach enhances the reliability of demand estimations under extreme temperature conditions.

\indent  An essential factor in extreme temperature analysis would be the cumulative temperature effect, wherein prolonged exposure to high or low temperatures significantly increases electricity demand for cooling or heating, even if the actual temperature remains unchanged. This phenomenon results in an apparent paradox where electricity demand continues to rise despite stable temperature levels, as extended exposure amplifies the need for indoor climate control. For instance, during a heatwave, electricity demand increases not only due to immediate high temperatures but also because prolonged exposure intensifies cooling requirements. In this context, the PDF, which captures only the relative frequency of temperatures without considering exposure duration, may exhibit weaker comovement with electricity demand, particularly under extreme conditions. Conversely, the LHR framework, which inherently accounts for the persistence of extreme temperature states, is expected to provide a more robust methodological foundation for assessing the dynamic relationship between extreme temperature events and electricity demand.

\indent It might be of interest to consider quantile function  corresponding to $\phi_t$ as the distributional predictor in our DSR model. The model with the quantile function predictor shares a conceptual similarity with the density-to-demand and hazard-to-demand models in that all three utilize the distributional properties of temperature as predictive variables. However, these models differ fundamentally in their interpretation of the demand response function \( f \). In the density-to-demand and hazard-to-demand models, the kernel-weighted benchmark response function \( \widetilde{\psi}(s) \) measures the impact of more frequently occurring temperatures around a given temperature level \( s \in [-20,40] \), either unconditionally or conditionally, while allowing for potential variations in associated temperature quantiles. In contrast, a similar quantity, which can similarly be defined in the quantile-to-demand model, represents the effect of changes around the \( s \)-quantile, where \( s \in [0,1] \), with the corresponding temperature level varying due to monthly (seasonal) fluctuations. 

\begin{figure}[t]
\begin{center}
\includegraphics[height=0.33\textwidth, width=0.5\textwidth]{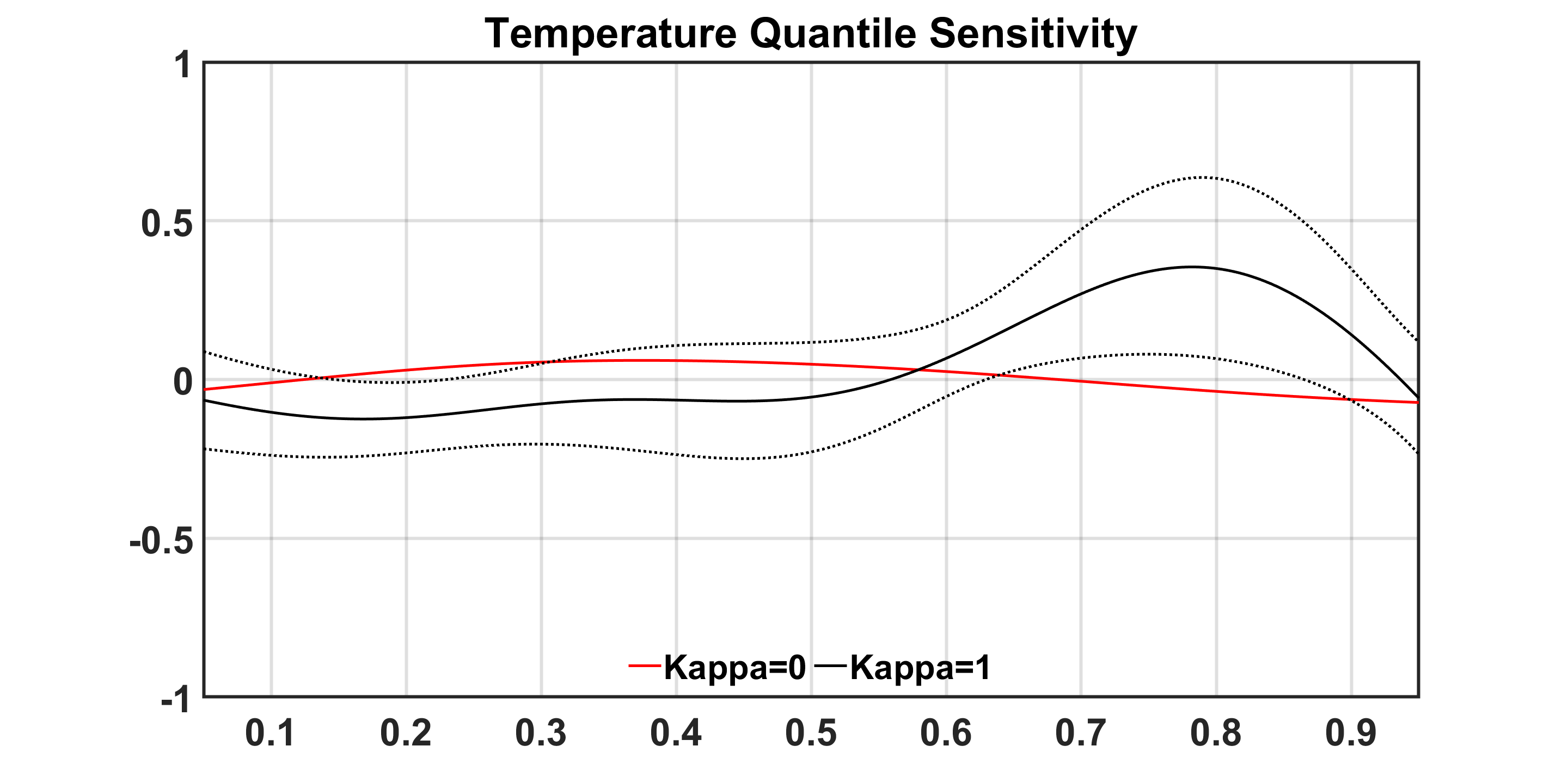}
\caption{Estimated quantile-based benchmark response function, truncated within the [0.05, 0.95] range, with dotted lines representing the 95\% confidence intervals.}
\label{Fig:Quantile_Descrip}
\end{center}
\end{figure}


To illustrate the difference between the quantile-to-demand model and the previously considered models, we additionally examine the impact of temperature quantiles on electricity demand during extreme weather events by estimating the quantile-to-demand model, a more detailed description of which is provided in Section \ref{app_quantile} of the Appendix.  Notably, the benchmark response function from the quantile-to-demand model, reported in Figure \ref{Fig:Quantile_Descrip}, remains largely insignificant, except within the 0.63 to 0.87 quantile range for $\kappa=1$. This contrasts with the temperature density and LHR sensitivity presented in the left panels of Figures \ref{Fig:Empirical_rlts} and \ref{Fig:HtoD_rlts}. The temperature quantile time series exhibits a monthly pattern, with lower values in winter and higher values in summer, whereas residential electricity demand exhibits a dual-peak structure, rising sharply in both summer and winter. This divergence introduces nonlinearities in the relationship between temperature and electricity demand, as an increase in temperature quantile time series simultaneously reduces heating demand while increasing cooling demand. Consequently, unlike an increase in the temperature density or LHR time series, which is unambiguously associated with higher cooling (heating) demand around 40°C (-20°C), applying fixed-quantile shocks in the quantile-to-demand framework would result in ambiguous slope estimates due to these offsetting effects. Similarly, incorporating the distributional characteristics into the quantile shocks in Figure \ref{Fig:Quantile_Shock} would not change the result of producing insignificant or ambiguous temperature sensitivity.

\bigskip
\section{Concluding Remarks}\label{Sec_conclude}
\noindent In this paper, we study the nonlinear temperature sensitivity of residential electricity demand using distributional predictors of temperature, which are likely to contain measurement errors. Given that residential electricity demand is highly sensitive to temperature fluctuations, accurately capturing these nonlinearities is critical, especially during extreme temperature events where underestimation could jeopardize grid stability. Additionally, measurement errors may stem from the assumption that nonparametrically estimated temperature distributions represent true distributions, compounded by data quality issues and limitations within the estimation process.

\indent We address these challenges by applying our DSR framework with distributional predictors. This approach allows for the effective identification of nonlinearities in the relationship between temperature distribution and electricity demand, even in the presence of measurement errors. Through the analysis of historical heat wave and cold wave shocks of varying intensities within a distributional context, we find that the hazard-to-demand model more reliably captures nonlinear temperature sensitivity than the density-to-demand model, particularly in extreme conditions. Our results indicate that residential electricity demand is significantly more sensitive to intense cold waves, with demand increasing sharply as temperatures decline. However, as cold wave severity intensifies, the rate of electricity demand growth diminishes, whereas for heat waves, demand growth accelerates with increasing intensity. 

\indent Given our findings, energy practitioners should prioritize incorporating detailed temperature distribution analysis, particularly when planning for extreme weather events. Our empirical findings suggest that residential electricity demand exhibits greater sensitivity to cold waves compared to heat waves, albeit with a slower rate of increase in demand, highlighting the need for advanced demand estimating models that account for these nonlinear sensitivities. Moreover, it is crucial to address potential measurement errors by adopting robust econometric techniques, in order to ensure that demand response strategies are based on accurate and reliable estimates. Our econometric approach would enhance grid stability and support effective energy management during extreme temperature events.

\bigskip

\bibliographystyle{elsarticle-harv} 
\bibliography{FRKETQ_P3/FRKET_biblio}

\bigskip
\appendix\label{Sec_Append}
\newpage 
\section{Technical Appendix}\label{sec_app_est}
This technical appendix provides a detailed discussion of the asymptotic properties of the proposed estimator along with the necessary mathematical preliminaries.
\subsection{Mathematical preliminaries} \label{sec_app_prelim}
This section briefly reviews essential mathematical concepts for the proposed estimator and introduce notation. 

Let $\mathcal H_1$ be a separable Hilbert space with inner product $\langle \cdot,\cdot \rangle_1$ and norm $\|\cdot\|_1$. A $\mathcal H_1$-valued random variable $Z$ is defined as a measurable map from the underlying probability space to $\mathcal H_1$. Such a random element $Z$ is said to be square-integrable if $\mathbb{E}[\|Z\|_1^2]<\infty$. Let $\mathcal H_2$ be another such Hilbert space; in the context of our DSR model, a relevant example of $\mathcal H_2$ is either $\mathcal H_1$  or $\mathbb{R}$ as Euclidean 1-space.  We let $\otimes$ denote the tensor product associated with $\mathcal H_1$ and $\mathcal H_2$; more specifically, for any $\zeta_1 \in \mathcal H_1$ and $\zeta_2 \in \mathcal H_2$, 
\begin{equation*}
\zeta_1\otimes \zeta_2 (\cdot) = \langle \zeta_1,\cdot \rangle_1 \zeta_2; 
\end{equation*}
note that $\zeta_1\otimes \zeta_2$ is a map from $\mathcal H_1$ to $\mathcal H_2$. For any mean-zero square-integrable random elements $Z_1$ and $Z_2$, $\mathbb{E}[Z_1 \otimes Z_2]$, defined as a map from $\mathcal H_1$ to $\mathcal H_2$, is called the (cross-)covariance of $Z_1$ and $Z_2$. We call $A:\mathcal H_1 \mapsto \mathcal H_2$ a bounded linear map if $A$ is a linear map and its operator norm, defined as $\|A\|_{\op}=\sup_{\|\zeta\|_1\leq 1} \|A\zeta\|_2$, is finite. For any bounded linear map $A$, we let $A^\ast$ be the adjoint map defined by the property $\langle A\zeta_1,\zeta_2 \rangle_2=\langle \zeta_1,A^\ast\zeta_2 \rangle_1$ for all $\zeta_1\in \mathcal H_1$ and $\zeta_2 \in \mathcal H_2$.  We say that $A$ is a compact operator if $\mathcal H_1=\mathcal H_2$ and there exist orthonormal bases $\{\zeta_{1,j}\}_{j\geq 1}$ and $\{\zeta_{2,j}\}_{j\geq 1}$ such that $A = \sum_{j=1}^\infty a_j \zeta_{1,j} \otimes \zeta_{2,j}$ with $a_j$ tending to zero as $j$ increases; in this expression, if $A$ is self-adjoint ($A=A^\ast$) and nonnegative ($\langle A\zeta,\zeta \rangle_1\geq 0$), it may be assumed that $a_j\geq 0$ and $\zeta_{1,j}=\zeta_{2,j}$, and furthermore they are understood as the eigenvalues and the corresponding eigenvectors. A compact operator $A$ is Hilbert-Schmidt if $\sum_{j=1}^\infty \|A \zeta_j\|_1^2 < \infty$ for some orthonormal basis $\{\zeta_j\}_{j\geq1}$, and in this case, the Hilbert-Schmidt norm $\|A\|_{\HS}$ is defined as  $\|A\|_{\HS}= \sqrt{\sum_{j=1}^\infty \|A \zeta_j\|_1^2}$. It is well known that $\|A\|_{\op} \leq \|A\|_{\HS}$ (see e.g., \citealp{Bosq2000}, (1.55)).

\subsection{Equivalent formulation of the proposed estimator} \label{sec_app_est_1}
Our estimator, discussed in Section \ref{Sec_Model}, is a suitable adaptation of the functional IV estimator proposed by \cite{seong2021functional} for our DSR model. First, observe that for an IV $Z_t$ satisfying \eqref{eqreg3}, the following can directly be deduced from \eqref{eqreg2}: $C_{Zy} = fC_{XZ}^\ast$, where $C_{Zy} = \mathbb{E}[(Z_t-\mathbb{E}[Z_t])\otimes (y_t-\mathbb{E}[y_t])]$. This in turn leads to the following identity:
\begin{align} \label{eqpopulation}
C_{Zy}C_{XZ} = f C_{XZ}^\ast C_{XZ},
\end{align}
where $C_{Zu}:=\mathbb{E}[(Z_t-\mathbb{E}[Z_t])\otimes u_t]=0$.
As in \cite{seong2021functional}, where the FFR model is studied, we construct an estimator that mimics the population relationship \eqref{eqpopulation}. We first replace $C_{Zy}$ and $C_{XZ}$ in \eqref{eqpopulation} with their sample counterparts $\widehat C_{Zy} =T^{-1}\sum_{t=1}^T (Z_t-\bar{Z}_T)\otimes (y_t-\bar{y}_T)$, where  $\bar{y}_T=T^{-1}\sum_{t=1}^T y_t$,  and $\widehat C_{XZ}$, which is introduced in Section \ref{Sec_Model}. Noting that $\widehat  C_{XZ}^\ast \widehat C_{XZ}$ is not invertible (see e.g., \citealp{Benatia2017}), we employ a regularized inverse of $\widehat  C_{XZ}^\ast \widehat C_{XZ}$ to compute an estimator from the sample equation. Specifically, 
we consider a rank-regularized inverse $(\widehat{C}_{XZ}^\ast \widehat{C}_{XZ})^\dag$ of $\widehat{C}_{XZ}^\ast \widehat{C}_{XZ}$ defined as follows:
\begin{equation} \label{eqregul}
(\widehat{C}_{XZ}^\ast \widehat{C}_{XZ})^\dag =  \sum_{j=1}^{\KK} {\hat{\lambda}_j^{-2}} \hat{g}_j \otimes \hat{g}_j,
\end{equation}
where, for $\alpha$ tending to zero, $\KK$ is determined as 
\begin{equation} \label{eqchoicek}
\KK = \max_{j\geq 1} \{\hat{\lambda}_j^2 \geq \alpha\}.
\end{equation}
Note that $\widehat{C}_{XZ}^\ast \widehat{C}_{XZ}(\widehat{C}_{XZ}^\ast \widehat{C}_{XZ})^\dag$  becomes the $\KK$-dimensional projection $\widehat{\Pi}_{\KK} = \sum_{j=1}^{\KK} \hat{g}_j \otimes \hat{g}_j$. Regularized inverses, similar to \eqref{eqregul}, are widely used in the literature on the functional linear model (see, e.g., \citealp{Chen_et_al_2020,seong2021functional,seo2024optimal}). The estimator $\hat{f}$ is constructed as follows: for any $x \in \mathcal H$
\begin{align} \label{eqest1}
\hat{f} (x)=  \widehat{C}_{Zy} \widehat{C}_{XZ} (\widehat{C}_{XZ}^\ast \widehat{C}_{XZ})^\dag (x).
\end{align}
Using \eqref{eqregul} and the definitions of the operators above, along with some algebra, it can be shown that \eqref{eqest1} is equivalent to the expression given in \eqref{eqest}.
\begin{remark} \label{rem1}
The choice $\KK$ in \eqref{eqchoicek} is adopted from \cite{seong2021functional} concerning the FFR model. Based on this choice, we study the asymptotic properties of our estimator. Of course, with only a minor modification, it is possible to extend the desired asymptotic results, which will be discussed, for an alternative choice of $\KK$, such as $\KK = m +  \max_{j\geq 1}\{\hat{\lambda}_j^2 \geq \alpha\}$ for some finite $m$. This choice may be preferred when researchers want to retain at least $m$ eigenelements for estimation. The case with $m=1$ also has some theoretical justification, given that we need at least one eigenelement for the proposed estimator to be well defined for every $T$.
\end{remark}

\subsection{Consistency and asymptotic normality}\label{sec_app_est_2}
This section establishes the consistency and asymptotic normality of the proposed estimator, both of which are fundamental to our empirical analysis in Section \ref{Sec_Empirics}. 

\subsubsection{Sone notational and theoretical simplifications}\label{sec_app_est2a}

For notational convenience, we hereafter let $\|\zeta_1\|$ denote the norm of $\zeta_1$ regardless of the Hilbert space in which $\zeta_1$ takes values. There is little risk of confusion following this simplification since, in the sequel, the considered Hilbert space is either $L^2[a,b]$ (the Hilbert space of square-integrable functions defined on $[a,b]$) or Euclidean 1-space $\mathbb{R}$. For any linear operator $A_T$ and a nonnegative real number $B_T$ depending on $T$, we let $A_T = O_p(B_T)$ (resp.\ $A_T = o_p(B_T)$) if $\|A_T\|_{\op}$ is $O_p(B_T)$ (resp.\ $o_p(B_T))$ whenever it is convenient as in \citet[Supplementary material]{seo2020functional}.

We subsequently consider the case where \( y_t \), \( X_t \), and \( Z_t \) have zero means, which reduces the regression model \eqref{eqreg1} to \( y_t = f(X_t^{\circ}) + \varepsilon_t \). Under this simplification, the population (e.g., \( C_{XZ} \)) and sample covariance operators (e.g., \( \widehat{C}_{XZ} \)) introduced in Section \ref{sec_method} are naturally interpreted as being defined without centering or demeaning the associated variables. As earlier discussed by \cite{seong2021functional}, extending the subsequent results to cases where the means are unknown and must be estimated requires only minor modifications, such as replacing variables with their centered or demeaned counterparts. Consequently, we omit these details.

\subsubsection{Assumptions and main results}
It turns out that $	C_{XZ}^\ast C_{XZ}$ and $C_{XZ} C_{XZ}^\ast $ are self-adjoint, nonnegative and compact (see \citealp{Bosq2000}, p.\ 117). We thus hereafter let 
\begin{equation*}
C_{XZ}^\ast C_{XZ} = \sum_{j=1}^\infty \lambda_j^2 g_j \otimes g_j, \quad C_{XZ} C_{XZ}^\ast = \sum_{j=1}^\infty \lambda_j^2 h_j \otimes h_j,
\end{equation*}
where $\lambda_j$, $g_j$, and $h_j$ are relevant eigenelements (see Section \ref{sec_app_prelim}). Let $C_{ZZ}=\mathbb{E}[Z_t\otimes Z_t]$, $\sigma_u^2=\mathbb{E}[u_t^2]$ and $C_{Zu}=\mathbb{E}[Z_t\otimes u_t]$. We let $\widehat{C}_{ZZ}$ and $\widehat{C}_{Zu}$ be defined as follows:
\begin{equation*}
\widehat{C}_{ZZ}=\frac{1}{T}\sum_{t=1}^T Z_t\otimes Z_t, \quad 
\widehat{C}_{Zu}=\frac{1}{T}\sum_{t=1}^T Z_t\otimes u_t,
\end{equation*}
which may be understood as the sample counterparts of $C_{ZZ}$ and $C_{Zu}$. We employ the following assumptions: below, $\mathfrak{F}_t$ denotes the natural filtration given by $\sigma(\{Z_s\}_{s\leq t+1}, \{u_s\}_{s\leq t})$ and let $\tau_j = 2\sqrt{2} \max\{ (\lambda_{j-1}^2-\lambda_j^2)^{-1}, (\lambda_{j}^2-\lambda_{j+1}^2)^{-1} \}$ as in \cite{seong2021functional}.
\begin{assumption}\label{assum1}
\begin{enumerate*}[(i)]
\item Equation \eqref{eqreg2} holds; \item $\{X_t,Z_t\}_{t\geq 1}$ is stationary and geometrically strongly mixing, $\mathbb{E}[\|X_t\|^2] < \infty$, and $\mathbb{E}[\|Z_t\|^2] < \infty$; \item $\mathbb{E}[u_t|\mathfrak F_{t-1}]=0$,  $\mathbb{E}[u_t^2|\mathfrak F_{t-1}]=\sigma_u^2$, and, for some $\delta>0$, $\sup_{1\leq t\leq T}\mathbb{E}[\|u_t\|^{2+\delta}|\mathfrak F_{t-1}] < \infty$; \item $\sum_{j=1}^\infty \|f(b_j)\|^2 < \infty$ for some orthonormal basis $\{b_j\}_{j\geq 1}$ of $\mathcal H$; \item $\|\widehat{C}_{XZ}-{C}_{XZ}\|_{\HS}=O_p(T^{-1/2})$, $\|\widehat{C}_{ZZ}-{C}_{ZZ}\|_{\HS}=O_p(T^{-1/2})$ and $\|\widehat{C}_{Zu}-{C}_{Zu}\|_{\HS}=O_p(T^{-1/2})$; \item $\ker C_{XZ}=\{0\}$; \item \label{eq001a1}$\lambda_1^2 > \lambda_2^2 > \cdots > 0$; \item \label{eq001a2}$T^{-1}\alpha^{-1} \to 0$ and $T^{-1/2}\sum_{j=1}^{\KK} \tau_j \to_p 0$. 
\end{enumerate*}
\end{assumption}
The conditions given in Assumption \ref{assum1} are adapted from Assumptions M and E of \cite{seong2021functional} studying the FFR model, and similar assumptions have been employed in the literature on the FSR model (see e.g., \citealp{Hall2007}). Condition \ref{eq001a2} is a technical requirement that $\alpha$ decay at an appropriate rate depending on the eigenvalues $\{\lambda_j\}$ and $T$. This subtle requirement on $\alpha$ may be relaxed if some additional (but still standard in the literature on the functional linear model) assumptions on the decay rate of $\lambda_j^2-\lambda_{j+1}^2$ as a function of $j$ are employed. For example, if Assumption \ref{assum2}, which is to appear, is satisfied, then it is straightforward to see that the consistency result given in Proposition \ref{prop1} can be established with e.g., \( \alpha^{-1} = o(T^{1/4}) \), without requiring Assumption \ref{assum1}\ref{eq001a2} (see Remark \ref{remrate} in Section \ref{sec_math}).
We next present the consistency result, which elaborates on Proposition \ref{prop1} given in Section \ref{Sec_Model}:

\begin{proposition} \label{prop1a}
Under Assumptions \ref{assum1}, $\|\hat{f} - f\|_{\op} \to_p 0$.  
\end{proposition}

We next obtain the asymptotic normality result \eqref{eqlocal} for any specified $\zeta \in \mathcal H$. To this end, we employ the following assumptions: below, we let $\upsilon_{t}(j,\ell) = \langle X_t,g_j \rangle\langle Z_t,h_{\ell}\rangle -\mathbb{E}[\langle X_t,g_j \rangle\langle Z_t,h_{\ell}\rangle]$ for $j,\ell \geq 1$. 

\begin{assumption}\label{assum2} For some $c_\circ>0$, $\rho>2$, $\varsigma>1/2$, $\delta_{\zeta}>1/2$ and $m>1$, the following holds:
\begin{enumerate*}[(i)]
\item $\lambda_j^2 \leq c_\circ j^{-\rho}$; \item \label{assum2b} $\lambda_j^2-\lambda_{j+1}^2 \geq c_\circ j^{-\rho-1}$; \item \label{assum2b2}  $\|f(g_j)\| \leq c_\circ j^{-\varsigma}$; \item \label{assum2c} $\mathbb{E}[\upsilon_t(j,\ell)\upsilon_{t-s}(j,\ell)] \leq c_\circ s^{-m}\mathbb{E}[\upsilon_t^2(j,\ell)]$ for $s\geq 1$, and furthermore, $\mathbb{E}[\langle X_t,g_j \rangle^4]\leq c_{\circ} \lambda_j^2$ and $\mathbb{E}[\langle Z_t,h_j \rangle^4]\leq c_{\circ} \lambda_j^2$; \item \label{assum2c2} $\langle g_j,\zeta \rangle \leq c_\circ j^{-\delta_{\zeta}}$.
\end{enumerate*}
\end{assumption}
The above assumptions are obvious adaptation of the conditions given by \cite{Hall2007} and \cite{seong2021functional}, which seem to be standard in the literature. Particularly, \cite{seong2021functional} established useful asymptotic results for the FFR model under assumptions similar to Assumption \ref{assum2}, and some of these asymptotic results are used to study the asymptotic properties of our estimator. In order to establish the asymptotic normality result, we need more conditions, which are stated below:
\begin{assumption}\label{assum3}
\begin{enumerate*}[(i)]
\item $\alpha^{-1} = o(T^{1/3})$;  \item \label{assum3b} $\varsigma + \delta_{\zeta} > \rho/2+2$, $2\delta_{\zeta} \geq \rho+1$ and $T \alpha^{(2\varsigma+2\delta_{\zeta}-1)/\rho} = O(1)$; \item\label{assum3c} ${\theta}_{\KK}(\zeta)= \langle  {\zeta}, ({C}_{XZ}^\ast {C}_{XZ})^\dag{C}_{XZ}^\ast {C}_{ZZ} {C}_{XZ} ({C}_{XZ}^\ast {C}_{XZ})^\dag {\zeta} \rangle \to_p \infty$.
\end{enumerate*}
\end{assumption} Assumption \ref{assum3}\ref{assum3b} requires $f$ and $\zeta$ to be sufficiently smooth with respect to the eigenvectors $\{g_j\}_{j\geq 1}$ so that the entire map or element can be well approximated by those eigenvectors; this is a standard requirement in asymptotic analysis in the literature on the functional linear model (see e.g., \citealp{seong2021functional} and reference therein). Assumption \ref{assum3}\ref{assum3c} requires a certain quantity depending on $\zeta$ to diverge to infinity. This condition is expected to be satisfied for many possible choices of $\zeta$ (see Section 3.2 of \citealp{Carrasco2007}; Remark 4 of \citealp{seong2021functional}).  Although the desired asymptotic normality result is established under these requirements, a weaker version of \eqref{eqlocal}, with \( f(\zeta) \) replaced by \( f\widehat{\Pi}_{\KK}(\zeta) \), still holds without requiring that these conditions be satisfied, as long as \( \alpha \) decays at a sufficiently slow rate. This allows for statistical inference on $f\widehat{\Pi}_{\KK}(\zeta)$. As discussed by \citet[Section 3.2]{seong2021functional}, $\widehat{\Pi}_{\KK}(\zeta)$ may be understood as the best linear approximation of $\zeta$ based on the covariation of $X_t$ and $Z_t$, thus this weaker result still provides interpretable insights for practitioners.

The desired asymptotic normality result is given as follows:
\begin{proposition} \label{prop2} Under Assumptions \ref{assum1}-\ref{assum3}
\begin{equation*}
\sqrt{\frac{T}{\hat{\sigma}^2_u\widehat{\theta}_{\KK}(\zeta)}}(\hat{f}(\zeta)-f(\zeta)) \to_d N(0,1),
\end{equation*}
where $\hat{\sigma}^2_u=T^{-1}\sum_{t=1}^T (y_t-\hat{f}(X_t))^2$ and  $\widehat{\theta}_{\KK}(\zeta)= \langle  {\zeta}, (\widehat{C}_{XZ}^\ast \widehat{C}_{XZ})^\dag \widehat{C}_{XZ}^\ast \widehat{C}_{ZZ} \widehat{C}_{XZ} (\widehat{C}_{XZ}^\ast \widehat{C}_{XZ})^\dag {\zeta} \rangle$.
\end{proposition}
With some algebra, it can be shown that $\widehat{\theta}_{\KK}(\zeta)$, as given in Proposition \ref{prop2} can be written as 
\begin{equation*}
\widehat{\theta}_{\KK}(\zeta) 
=\frac{1}{T} \sum_{t=1}^T \left(\sum_{j=1}^{\KK} \hat{\lambda}_j^{-2} \langle \hat{g}_j, \zeta\rangle \langle \widehat{C}_{XZ}\hat{g}_j, Z_t\rangle\right)^2,
\end{equation*}
which corresponds to \eqref{eqtheta} without demeaning the variables.

\subsubsection{Mathematical proofs} \label{sec_math}
\subsubsection*{Proof of Proposition \ref{prop1a}}
Using \eqref{eqreg2} and the expression given in \eqref{eqest1}, the proposed estimator $\hat{f}$ can be written as 
\begin{equation}
\hat{f}  =   \widehat{C}_{Zy} \widehat{C}_{XZ} (\widehat{C}_{XZ}^\ast \widehat{C}_{XZ})^\dag	=  f \widehat{\Pi}_{\KK} +  \widehat{C}_{Zu} \widehat{C}_{XZ} (\widehat{C}_{XZ}^\ast \widehat{C}_{XZ})^\dag, \label{eq:b0}
\end{equation}
where $\widehat \Pi_{\KK}= \sum_{j=1} ^{\KK} \widehat g_j \otimes \widehat g_j$ as earlier introduced.  
As shown in the proof of Theorem 1 of \cite{seong2021functional}, we have  $\|\widehat{C}_{XZ} (\widehat{C}_{XZ}^\ast \widehat{C}_{XZ})^\dag\|_{\op} \leq \alpha^{-1/2}$ and $\|\widehat{\mathcal C}_{Zu}\|_{\HS}=O_p(T^{-1/2})$ under Assumption \ref{assum1}. Consequently, we find that
\begin{equation*}
\|\widehat{f} - f\widehat{\Pi}_{\KK}\|_{\op} \leq \|\widehat{\mathcal C}_{Zu}\|_{\HS}  \|\widehat{C}_{XZ} (\widehat{C}_{XZ}^\ast \widehat{C}_{XZ})^\dag\|_{\op}  \leq O_p(\alpha^{-1/2} T^{-1/2}).
\end{equation*}
We hereafter assume that $\{\hat{g}_j\}_{j\geq 1}$ consists of an orthonormal basis; this can simply be achieved by assigning an appropriate vector to each zero eigenvalue with no loss of generality (see e.g., the proof of Lemma A.1 of \citealp{seo2024optimal}). From the properties of the operator norm and the fact that $\{\hat{g}_j\}_{j\geq 1}$ is an orthonormal basis, we find the following:
\begin{equation*}
\|f\widehat{\Pi}_{\KK} -f\|^2_{\op} \leq \sum_{j=1}^\infty \|f\widehat{\Pi}_{\KK}(\hat{g}_j) -f\hat({g}_j)\|^2. 
\end{equation*}
Thus, the proof becomes complete if $\sum_{j=1}^{\infty}\|f\widehat{\Pi}_{\KK}(\hat{g}_j)-f(\hat{g}_j)\|^2 \to_p 0$ is shown. From  
nearly identical arguments used to derive (8.63) of \cite{Bosq2000}, 
we find that 
\begin{equation}
\sum_{j=1}^{\infty} \|f\widehat{\Pi}_{\KK}(\hat{g}_j) -f(\hat{g}_j)\|^2  \leq 	\sum_{j=\KK+1}^{\infty} \|f(\hat{g}_j)\|^2 \leq 	\sum_{j=\KK+1}^\infty \|f({g}_j)\|^2 + O_p(1)\|\widehat{ C}_{XZ}^\ast  \widehat{ C}_{XZ}- C_{XZ}^{\ast} C_{XZ}\|_{\op} \sum_{j=1}^{\KK} \tau_j.\label{eqpf0006add2wadd}
\end{equation}
Since $\sum_{j=\KK+1}^\infty \|f(g_j)\|^2 < \infty$ under Assumption \ref{assum1}, $\sum_{j=\KK+1}^\infty \| f(g_j)|^2$ converges in probability to zero as $T$ gets larger (note that $\KK$ diverges as $T \to \infty$). In addition, 
$
\|\widehat{ C}_{XZ}^\ast  \widehat{ C}_{XZ}- C_{XZ}^{\ast} C_{XZ}\|_{\op} \leq \|\widehat{ C}_{XZ}^\ast\|_{\op}\|\widehat{ C}_{XZ}- C_{XZ}\|_{\op} + \|\widehat{ C}_{XZ}^\ast- C_{XZ}^\ast\|_{\op}\| C_{XZ}\|_{\op} = O_p(T^{-1/2})$ by the properties of the operator norm and Assumption \ref{assum1}. Combining all these results, we find that  \eqref{eqpf0006add2wadd} is $o_p(1)$ as desired.  \qed

\subsubsection*{Proof of Proposition \ref{prop2}}
In this proof, we let $g_j^s = \sgn(\langle \hat{g}_j, g_j \rangle)g_j$ as in the proof of Theorem 2 of \cite{seong2021functional}. We also note that $\widehat{\theta}_{\KK}$ can equivalently 
be written as $\widehat{\theta}_{\KK}(\zeta)= \langle  {\zeta}, (\widehat{C}_{XZ}^\ast \widehat{C}_{XZ})^\dag\widehat{C}_{XZ}^\ast \widehat{C}_{ZZ} \widehat{C}_{XZ} (\widehat{C}_{XZ}^\ast \widehat{C}_{XZ})^\dag {\zeta} \rangle.$

From nearly identical arguments used in the proof of Theorem 2 of \cite{seong2021functional} (particularly, see (S2.4)-(S2.6) and (S2.11) of their paper) and Assumptions \ref{assum1}-\ref{assum2}, it can be shown that $\alpha \KK^\rho  = O_p(1)$ and also
\begin{equation} \label{eqadda0}
\| \widehat{C}_{XZ} (\widehat{C}_{XZ}^\ast \widehat{C}_{XZ})^\dag -  {C}_{XZ} ({C}_{XZ}^\ast {C}_{XZ})^\dag\|_{\op} \leq O_p(\alpha^{-1/2}T^{-1/2} \sum_{j=1}^{\KK}\tau_j) \leq O_p(T^{-1/2} \alpha^{-3/2-2/\rho}) = o_p(1).
\end{equation}
We observe that 
\begin{equation} \label{eqdecom}
\hat{f}(\zeta) -f(\zeta) = (\hat{f}-f\widehat{\Pi}_{\KK})(\zeta) + f(\widehat{\Pi}_{\KK}-\Pi_{\KK})(\zeta) + f(\Pi_{\KK}-I)(\zeta),
\end{equation}
where $\Pi_{\KK} = \sum_{j=1}^{\KK} g_j^s \otimes g_j^s$. The first term in \eqref{eqdecom}, $(\hat{f}-f\widehat{\Pi}_{\KK})(\zeta)$, satisfies the following: 
\begin{align} \label{eqconv1}
\sqrt{\frac{T}{{\sigma}^2_u{\theta}_{\KK}(\zeta)}} (\hat{f}-f\widehat{\Pi}_{\KK})(\zeta)
&=\sqrt{\frac{1}{{\sigma}^2_u{\theta}_{\KK}(\zeta)}} \left(\frac{1}{\sqrt{T}}\sum_{t=1}^T Z_t \otimes u_t\right) \left({C}_{XZ} ({C}_{XZ}^\ast {C}_{XZ})^\dag + o_p(1) \right) (\zeta),
\end{align}
where the equality follows from simple algebra (as in \eqref{eq:b0}) and \eqref{eqadda0}. If define $\xi_t = {\sigma_u^{-1}{\theta}_{\KK}(\zeta)}^{-1/2} (Z_t \otimes u_t) {C}_{XZ} ({C}_{XZ}^\ast {C}_{XZ})^\dag (\zeta)$, then $\sigma_u^{-1} \theta_{\KK}(\zeta)^{-1/2} \left(\frac{1}{\sqrt{T}}\sum_{t=1}^T Z_t \otimes u_t\right) {C}_{XZ} ({C}_{XZ}^\ast {C}_{XZ})^\dag (\zeta) = T^{-1/2}\sum_{t=1}^T \xi_t$. Then as can obviously be deduced from (S2.7) and (S2.8) by \cite{seong2021functional}, the above $\{\xi_t\}$ is a martingale difference sequence (w.r.t. $\mathcal F_{t-1})$ and also $\mathbb{E}[\xi_t^2] = 1$. Then, from the employed assumption and the standard central limit theorem for a martingale difference sequence, we find that $ T^{-1/2}\sum_{t=1}^T \xi_t  \to_d N(0,1)$, and hence we conclude that  
\begin{equation*}
\sqrt{\frac{T}{{\sigma}^2_u{\theta}_{\KK}(\zeta)}} (\hat{f}(\zeta)-f\widehat{\Pi}_{\KK}(\zeta))  \to_d N(0,1). 
\end{equation*}

We then note that the second term in \eqref{eqdecom}, $f(\widehat{\Pi}_{\KK}-\Pi_{\KK})(\zeta)$, can be rewritten as  $\sqrt{T/\theta_{\KK}(\zeta)} (A_1+A_2+A_3)$, where $A_1 = \sum_{j=1}^{\KK} \langle \hat{g}_j-g_j^s,\zeta \rangle f(\hat{g}_j-g_j^s)$, $A_2 =\sum_{j=1}^{\KK} \langle g_j^s,\zeta \rangle f(\hat{g}_j-g_j^s)$,  $A_3 =\sum_{j=1}^{\KK} \langle \hat{g}_j-g_j^s,\zeta \rangle f(g_j)$. To analyze these terms, we first investigate the stochastic order of the quantity $\langle (\hat{C}_{XZ} -{C_{XZ}}) g_j,h_{\ell}\rangle$. From Assumption \ref{assum2} and nearly identical arguments used in the unnumbered equation between (S2.28) and (S2.29) in \cite{seong2021functional}, the following can be shown: 
\begin{equation} \label{eqadd01a}
T\mathbb{E}[\langle (\hat{C}_{XZ} -{C_{XZ}}) g_j,h_{\ell}\rangle^2] \leq \sum_{s=0}^T\mathbb{E}[v_t(j,\ell)v_{t-s}(j,\ell)]\leq O(1) \mathbb{E}[\langle X_t,g_j \rangle^2\langle Z_{t},h_{\ell} \rangle^2] = O(\lambda_j\lambda_{\ell}), 
\end{equation} where the second inequality follows from Assumption \ref{assum2}\ref{assum2c} and the third inequality follows from the Cauchy-Schwarz inequality and Assumption \ref{assum2}\ref{assum2c}. \eqref{eqadd01a} implies that $\langle (\hat{C}_{XZ} -{C_{XZ}}) g_j,h_{\ell}\rangle=O_p(T^{-1/2}\sqrt{\lambda_j \lambda_{\ell}})$. Combining this result with Lemma S1 of \cite{seong2021functional}, we conclude that $\|\hat{v}_j-v_j\|^2 = O_p(T^{-1}j^2)$ holds under Assumption \ref{assum1}-\ref{assum3}. In turn, from similar arguments used in the proofs of (S2.16) and (S2.33)  in \cite{seong2021functional}, the following can also be deduced:
\begin{align*}
\|f(\hat{g}_j-g_j^s)\|^2 = O_p(T^{-1}) j^{\rho+2-2\varsigma},\\
\langle \hat{g}_j-g_j^s,\zeta \rangle^2 = O_p(T^{-1}) j^{\rho+2-2\delta_{\zeta}}.
\end{align*} 
Combining these results with the conditions given in Assumption \ref{assum3}, it can be shown (see pp.\ S12-S13 of the Supplementary Material in \citealp{seong2021functional}) that 
\begin{equation} \label{eqconv1a}
\sqrt{T}(\|A_1\| + \|A_2\| + \|A_3\|) \leq   O_p(\sum_{j=1}^{\KK} j^{\rho/2 - \varsigma-\delta + 1}) = O_p(\KK^{\rho/2 - \varsigma-\delta + 2}) = o_p(1), 
\end{equation}
where the last equality is deduced from that  $\KK \to \infty$ as $T\to \infty$ and $\rho/2 - \varsigma-\delta + 2 < 0$ under Assumption \ref{assum3}.  

The third term in \eqref{eqdecom}, $f({\Pi}_{\KK}-I)(\zeta)$, satisfies that 
\begin{equation}\label{eqconv1aa}
\|f({\Pi}_{\KK}-I)(\zeta)\|^2\leq \sum_{j=\KK+1}^{\infty} \|\langle g_j^s,\zeta \rangle f(g_j^s) \|^2 
\leq  O(1)\sum_{j=\KK+1}^{\infty} j^{-2\delta_{\zeta}-2\varsigma} =  O(\KK^{-2\delta_{\zeta}-2\varsigma+1})= O(\alpha^{(2\varsigma+2\delta_{\zeta}-1)/\rho}),
\end{equation}
where the second inequality follows from Assumptions \ref{assum2}\ref{assum2b2} and \ref{assum2}\ref{assum2c2},  the first equality follows from the Euler-Maclaurin summation formula for the Riemann zert-function (see, e.g., (5.6) of \citealp{ibukiyama2014euler}) and the second equality follows from the fact that $\alpha \KK^\rho  = O_p(1)$.  
Therefore, from \eqref{eqconv1a} and \eqref{eqconv1aa}, we find that
\begin{equation*}
\sqrt{T/\theta_{\KK}(\zeta)} ({f}\widehat{\Pi}_{\KK}-f)(\zeta) = O_p(1/\sqrt{\theta_{\KK}(\zeta)}) + O(T^{1/2}\alpha^{(\varsigma+\delta_{\zeta}-1/2)/\rho} /\sqrt{\theta_{\KK}(\zeta)}).
\end{equation*}
Under Assumption \ref{assum3}, the above is $o_p(1)$. We therefore conclude that 
\begin{equation*}
\sqrt{\frac{T}{{\sigma}^2_u{\theta}_{\KK}(\zeta)}} (\hat{f}-f)(\zeta) = \sqrt{\frac{T}{{\sigma}^2_u{\theta}_{\KK}(\zeta)}} (\hat{f}-f\widehat{\Pi}_{\KK}(\zeta)) + o_p(1) \to_d N(0,1).
\end{equation*}
To establish the desired asymptotic normality result, it only remains to show that (i) $\|\hat{\sigma}_u^2- {\sigma}_u^2\| \to_p 0$ and (ii) $\|\widehat{\theta}_{\KK}(\zeta)-\theta_{\KK}(\zeta)\| \to_p 0$. Under the employed assumptions, the former follows directly from the consistency of $\hat{f}$. The latter follows from \eqref{eqadda0}, as discussed by \cite{seong2021functional} in their proof of Theorem 2. \qed 



\begin{remark} \label{remrate}
As shown in the earlier proof of Proposition \ref{prop1a}, under Assumption \ref{assum1}, we have $\|\hat{f}-f\|_{\op} = O_p(\alpha^{-1/2}T^{-1/2})  + O_p(\sum_{j=\KK+1}^\infty \|f(g_j)\|^2) + O_p(T^{-1/2} \sum_{j=1}^{\KK} \tau_j)$, where the first two terms are $o_p(1)$ as shown in our earlier proof. Suppose further that $\alpha = o(T^{1/4})$ and Assumption \ref{assum2} are satisfied. Then as shown by \cite{seong2021functional}, $\alpha \KK^{\rho} = O_p(1)$, and by combining this result with Assumption \ref{assum2}, we find that 
$O_p(T^{-1/2} \sum_{j=1}^{\KK} \tau_j) \leq O_p(T^{-1/2}\sum_{t=1}^{\KK} j^{\rho+1}) = O_p(T^{-1/2} \KK^{\rho+2}) = O_p(T^{-1/2}\alpha^{-(\rho+2)/\rho})$, which is $o_p(1)$ since $\alpha=o(T^{1/4})$ and $\rho>2$.
\end{remark}

\section{Quantile function predictor} \label{app_quantile}
In Section \ref{sec_more}, we considered the quantile-to-demand model for comparison with the density-to-demand and hazard-to-demand models. In the DSR model given in \eqref{eqreg1}, as an alternative to the CLR, LHR, or LRHR predictors, practitioners may consider the quantile function corresponding to $\phi_t$ as the distributional predictor $X_t^\circ$. In this case,
\begin{equation*} 
X_t^\circ(r) = \inf \left\{ x \in [a,b] : \int_{a}^{x} \phi_t(s) \, ds \leq r \right\}, \quad r \in [a,b].  
\end{equation*}  
Like other distributional predictors, the quantile function predictor must be estimated in advance from raw temperature data. This can be done by replacing $\phi_t$ with the estimate $\hat{\phi}_t$ as described in Section \ref{Sec_Data_1}. Alternatively, the quantile function predictor can directly be estimated from the raw temperature data by replacing population quantiles with sample quantiles. Specifically, if \( x_{[j]} \) for \( j=1,\ldots,N_t \) are the ordered (smallest to largest) raw temperature observations at time \( t \), we may define  
\begin{equation} \label{qfpredictor}  
X_t(r) = x_{[N_t r]} \quad \text{if \( N_t r \) is an integer,}  
\end{equation}  
for \( r \in [0,1] \). If \( N_t r \) is not an integer, we employ a linear interpolation scheme to estimate values proportionally between neighboring data points. The quantile function predictor $X_t$ constructed in this way may be regarded as a reasonable proxy for $X_t^\circ$. Thus, we estimate the model in \eqref{eqreg2} as in the other DSR models considered in Sections \ref{DtoD_Rlt} and \ref{DtoD_Rlt2}. The results reported in Section \ref{sec_more} are based on the quantile predictor constructed as in \eqref{qfpredictor}.

\end{document}